\documentclass[pdftex,twocolumn,epjc3]{svjour3}          
\usepackage{subfig}
\RequirePackage[T1]{fontenc}
\usepackage{amsmath}
\usepackage{amssymb}
\usepackage{booktabs}
\usepackage{siunitx}  
\smartqed  

\RequirePackage{graphicx}
\RequirePackage{flushend}
\RequirePackage[numbers,sort&compress]{natbib}
\RequirePackage[colorlinks,citecolor=blue,urlcolor=blue,linkcolor=blue]{hyperref}
 
\def\be{\begin{equation}}
\def\ee#1{\label{#1}\end{equation}}
\newcommand{\ben}{\begin{eqnarray}}
\newcommand{\een}{\end{eqnarray}}

\begin{document}

\title{Shadows and amplitude luminosity of an embedded rotating black hole} 
\author{Abra\~ao J. S. Capistrano\thanksref{e1,addr1,addr2}
        \and
        Antonio C. Gutiérrez-Pineres\thanksref{e2,addr3,addr4} 
        \and 
        Carlos H. Coimbra-Ara\'ujo\thanksref{e3,addr1,addr2}
}

\thankstext{e1}{e-mail:capistrano@ufpr.br}
\thankstext{e2}{e-mail:acgutier@uis.edu.co}
\thankstext{e3}{e-mail:carlos.coimbra@ufpr.br}

\institute{Departamento de Engenharias e Ci\^encias Exatas, Universidade Federal do Paran\'a, Rua Pioneiro, 2153, Palotina/PR, Brazil\label{addr1}
          \and
           Graduate Program in Applied Physics, Federal University of Latin-American Integration, Avenida Tarqu\'inio Joslin dos Santos, 1000 - Polo Universit\'ario, Foz do Igua\c{c}u/PR, Brazil\label{addr2}
          \and
          Escuela de Fısica, Universidad Industrial de Santander, A. A. 678, Bucaramanga 680002, Colombia\label{addr3}
          \and
          Instituto de Ciencias Nucleares, Universidad Nacional Autónoma de México, AP 70543, DF 04510, Mexico\label{addr4}    
}
\date{Received: date / Accepted: date}

\maketitle

\begin{abstract}
We investigate a rotating black hole embedded in a five-dimensional flat bulk by extending the Kerr metric through the Gürses–Gürsey line element. Employing Boyer–Lindquist coordinates, we reinterpret the black hole as a charged-like object in five dimensions and analyze its horizon structure and shadow morphology. Our results reveal that the shadow is shaped by an axially symmetric gravitational field modulated by an extrinsic curvature term arising from the higher-dimensional embedding. Simulations demonstrate that the visibility amplitude and shadow profile of the Gürses-Gürsey black hole align with Event Horizon Telescope observations of M87$^{\ast}$ in the Kerr limit, while also allowing measurable deviations that could be probed by future high-resolution experiments.
\end{abstract}

\section{Introduction}
Black holes (BHs) are among the most fascinating and mysterious entities in the cosmos. Predicted by general relativity (GR), their existence has been supported by a range of observational techniques, including optical and X-ray measurements as well as gravitational wave detections. Despite compelling astrophysical evidence, questions remain as to whether the observed compact objects fully conform to the BHs described by GR, or if they might instead reflect predictions from alternative theories of gravity.
Astrophysical BHs appear to exhibit key features of rotating GR BHs, most notably, mass and spin
\cite{batal,bambi2013rotating,reynolds2021observational}. These characteristics are foundational to the Kerr solution \cite{kerr}, a mathematical model derived from Einstein's field equations that describes rotating BHs. The dynamics of particles and light around such objects are governed by the separability of the Hamilton-Jacobi equations, as shown by Carter, which introduces four conserved quantities including the Carter constant. Additionally, the effective potential framework--used to analyze the motion of test particles--provides insight into stable orbital configurations, particularly the innermost stable circular orbit (ISCO), which plays a critical role in accretion disk physics and observational signatures.

On the other hand, BHs are also expected to exert an extraordinary gravitational pull, strong enough to prevent even light from escaping. The surface beyond which no signal can reach a distant observer is known as the event horizon. This horizon marks the boundary between the black hole's interior and the external universe. As such, exploring the near-horizon region offers a valuable window into the nature of the interior. This can be approached by analyzing geodesic motion within the surrounding spacetime and by studying the distribution of light near the horizon—both of which provide critical insights into the geometry and physical behavior of BHs.

Indeed, in recent years, significant advances have been made in the detection and characterization of astrophysical BHs, positioning these observations as powerful laboratories for testing general relativity (GR). Evidence increasingly suggests that most galaxies harbor a supermassive black hole at their center, with some exhibiting active galactic nuclei (AGNs) that emit radiation across the entire electromagnetic spectrum. A landmark achievement in this domain was the Event Horizon Telescope (EHT) collaboration's imaging of two supermassive BHs--Sgr A$^{\ast}$ at the center of the Milky Way and M87$^{\ast}$ in the Virgo cluster. These observations represent a major milestone, offering unprecedented insight into the near-horizon structure and providing a direct test of GR in the strong-field regime
(see, e.g., \cite{akiyama,kocherlakota}). These observations not only affirm the predictions of GR regarding BHs in the strong-field regime, but also offer a compelling motivation to explore the shadow cast by BHs. The morphology and structure of these shadows provide a promising avenue for testing theoretical models, including those arising from alternative gravitational frameworks. As observational capabilities continue to advance, particularly in high-resolution imaging and interferometry, the near-horizon region and its associated light distribution may yield critical insights into the nature of gravity and spacetime.

In this paper, we develop a covariant and model-independent framework for gravitational perturbation theory grounded in Nash's embedding theorem~\cite{Nash1956}. Departing from conventional brane-world scenarios~\cite{Randall1999,Randall1999b,ArkaniHamed1998,Dvali2000,Battye2001,Davis2003,Yamauchi2007}, our approach treats the extrinsic curvature as an independent dynamical quantity—an orthogonal component of the gravitational field that complements the traditional tangent component described by the metric  $g_{\mu\nu}$. This embedding-based formulation offers a more general and flexible treatment of spacetime deformations, allowing us to explore a broader spectrum of perturbative phenomena while maintaining covariance throughout the model. Within this framework, we investigate black hole shadows and visibility amplitudes-key observational probes for testing GR and its extensions, especially in the strong-field regime near event horizons.

The black hole shadow, characterized by a dark central region encircled by a luminous emission ring, is a gravitationally lensed image of the photon capture zone. It encodes critical information about the underlying spacetime geometry, including the black hole's mass, spin, and possible deviations from the Kerr solution~\cite{Perlick2021,Akiyama2019}. Complementing this, the visibility amplitude—defined as the modulus of the Fourier transform of the black hole image obtained via very long baseline interferometry (VLBI)—reveals structural details of the source in the Fourier domain and remains notably resilient to calibration errors~\cite{Medeiros2016}. In the case of rotating BHs, features such as asymmetric ring brightness and nulls in the visibility amplitude are particularly sensitive to the inclination of the accretion disk and the geometry of the surrounding emission~\cite{Johnson2020, Chael2016}.

In higher-dimensional models—such as those inspired by string theory, including the Randall–Sundrum framework and its extensions—the morphology of black hole shadows can diverge markedly from four-dimensional predictions~\cite{Amarilla2012, Vagnozzi2022}. These extra-dimensional geometries often modify the properties of the photon sphere or produce multiple shadow features due to gravitational lensing by higher-dimensional structures~\cite{Nedkova2013}. However, the corresponding signatures in the visibility amplitude are typically subtle, demanding exceptionally high angular resolution and dense baseline coverage--capabilities that current Earth-based VLBI arrays only partially achieve~\cite{Akiyama2025}. Moreover, strong astrophysical effects, such as turbulent accretion flows and temporal variability, introduce degeneracies that can obscure or imitate the observational imprints of extra dimensions~\cite{Wielgus2022}.

Interpreting these observational signatures requires high-fidelity general relativistic magnetohydrodynamic (GRMHD) simulations adapted to higher-dimensional metrics, but it still remains computationally intensive~\cite{Mizuno2018}. Besides, there is absence of a model-independent framework capable of linking visibility amplitude features to specific extra-dimensional parameters, which constrains the direct empirical testing of such theories. Nevertheless, ongoing improvements on imaging algorithms and the anticipated deployment of space-based VLBI missions are poised to significantly enhance sensitivity to fine-scale deviations. These developments on visibility-based shadow analysis may be a compelling avenue for probing fundamental physics beyond the four-dimensional spacetime~\cite{Roelofs2019}.

The paper is organized as follows. In Section II, we present a summary of our model grounded in Nash's embedding theorem. Section III revisits results from previous work involving the Gürses–Gürsey metric~\cite{gursey}, embedded within a five-dimensional spacetime framework. In Section IV, we employ the Hamilton–Jacobi formalism to derive the integrals of motion, analyze the effective potential, and investigate the horizon structure, including the identification of the innermost stable circular orbit (ISCO). Section V investigates a cosmic-scale correction aimed at constraining the free parameter of our model using recent cosmological observations~\cite{Capistrano2024}. We also compute the modified Komar mass and analyze the associated energy conditions. In Section VI, we explore the shadow structures and visibility amplitudes of the spacetime, employing the Adaptive Analytical Ray Tracing (AART) code~\cite{low_lum_2024,Desire_2025}, which is publicly available at \url{https://github.com/iAART/aart}. For clarity, we outline key notation conventions: capital Latin indices range from 1 to 5; lowercase Latin indices denote the single extra dimension; Greek indices span the embedded four-dimensional spacetime. An overbar indicates a background (non-perturbed) quantity, and we adopt the Landau–Lifshitz sign convention for the metric signature. Finally, in the concluding section, we summarize our findings and discuss future directions for extending this framework.

\section{A summary of Nash embeddings}

To explore the physics of space-time embeddings, we adopt a simplified yet robust classical framework. Our analysis is primarily motivated by long-standing issues in fundamental physics, such as the hierarchy problem, as well as the persistent challenges posed by dark matter and dark energy problems. These phenomena suggest that the left-hand side (LHS) of Einstein's field equations might be incomplete or missing a key geometric contribution. One promising avenue to address this deficiency lies in embedding theories.

In this work, we adopt a dynamical embedding framework that stands apart from the more commonly studied braneworld models. In standard braneworld scenarios, perturbations arise from fields confined to the brane, often modeled by a scalar field coupled to gravity. These models typically rely on junction conditions \cite{Randall1999b} to replace the extrinsic curvature in terms of the energy-momentum content on the brane. However, this requires a complex mechanism to localize matter. In contrast, our approach adopts the extrinsic geometry as a dynamical degree of freedom for the gravitational field that ``leaks'' into the higher-dimensional space, whereas the gauge interactions remain confined in the embedded four-dimensional space. In fact, conventional braneworld models can be recovered as special cases within our covariant and geometrically independent construction.

The mathematical foundation for such isometric and differentiable embedding was originally introduced by J. Nash in 1956~\cite{Nash1956} and extended to Riemannian and pseudo-Riemannian geometries by Greene~\cite{greene}. Nash's result guarantees the possibility of embedding a lower-dimensional manifold $V_n$ into a higher-dimensional one $V_D$ (with $D > n$) via a differentiable map $\mathcal{X}: V_n \rightarrow V_D$, realized through a sequence of smooth metric deformations. The deformed metric takes the form
\begin{equation}\label{eq:York}
g_{\mu\nu}  =  \bar{g}_{\mu\nu}  +  \delta y \bar{k}_{\mu\nu}  +
(\delta y)^2 \bar{g}^{\rho\sigma} \bar{k}_{\mu\rho}\bar{k}_{\nu\sigma} + \cdots ,
\end{equation}
where $\bar{k}_{\mu\nu}$ denotes the extrinsic curvature, and $\delta y$ is a coordinate along the extra (orthogonal) dimension. The overbar indicates quantities associated with the background (non-perturbed) geometry. Using a Gaussian base $\{X^A_\mu, \bar{\eta}^A_a\}$, the extrinsic curvature is defined by
\begin{equation}
\bar{k}_{\mu\nu} =  -\mathcal{X}^A_{,\mu}\eta^B_{,\nu} \mathcal{G}_{AB}\;, \label{eq:extrinsic}
\end{equation}
where $\eta^A$ is a unit vector field normal to the embedded manifold, and $\mathcal{G}_{AB}$ is the bulk metric. This expression quantifies the bending of the embedded space within the higher-dimensional manifold and plays a central role in the geometric dynamics of the embedding.

Nash's deformation formula follows from Eq.\eqref{eq:York} and gives the relation:
\begin{equation}\label{eq:nashdeformation}
k_{\mu\nu} = -\frac{1}{2} \frac{\partial g_{\mu\nu}}{\partial y} \;,
\end{equation}
representing the rate of change of the induced metric with respect to the orthogonal direction. This expression encodes the essence of Nash's theorem and is a fundamental constraint on how perturbations propagate in the embedded geometry.

To trace these deformations, we use Lie transport of the Gaussian frame vectors, leading to the following relations
\begin{eqnarray}
\mathcal{Z}^A_{,\mu} &=& X^{A}_{,\mu} + \delta y\;\bar{\eta}^{A}_{,\mu}\;, \label{eq:pertu1} \\
\eta^A &=& \bar{\eta}^A + \delta y [\bar{\eta}, \bar{\eta}]^A = \bar{\eta}^A \;, \label{eq:pertu2}
\end{eqnarray}
where the perturbation vector $\eta^A$ remains invariant due to its independence from the embedded coordinates. This feature eliminates gauge ambiguities that could otherwise arise from coordinate-dependent perturbations. These relations ensure a well-defined and differentiable chart $\mathcal{X}: V_n \rightarrow V_D$, linking the embedded space to the bulk.

In our application, we consider a five-dimensional bulk $V_5$ with a four-dimensional embedded submanifold $V_4$. The corresponding metric can be written in a form
\begin{equation}
\mathcal{G}_{AB} = \begin{pmatrix}
g_{\mu\nu} & 0 \\
0 & 1
\end{pmatrix} \;, \label{eq:metricbulk}
\end{equation}
where $g_{\mu\nu}$ is the induced four-dimensional metric. The gravitational action defined over the bulk is:
\begin{equation}\label{eq:action}
S = -\frac{1}{2\kappa^2_5} \int \sqrt{|\mathcal{G}|} \left( {}^5\mathcal{R} + \mathcal{L}^{\*}_m \right) d^5x \;,
\end{equation}
where ${}^5\mathcal{R}$ is the five-dimensional Ricci scalar, $\kappa_5$ is the five-dimensional gravitational coupling, and $\mathcal{L}^{*}_m$ is the bulk matter Lagrangian.

Another key aspect of the embedding framework lies in the Gauss-Codazzi equations, which connect the intrinsic and extrinsic geometries, or, in other words, they relate the bulk with the embedded space and vice--versa. Such equations are given by 
\begin{eqnarray}
{}^5\mathcal{R}_{ABCD} \mathcal{Z}^A_{,\alpha} \mathcal{Z}^B_{,\beta} \mathcal{Z}^C_{,\gamma} \mathcal{Z}^D_{,\delta} &=& R_{\alpha\beta\gamma\delta} + (k_{\alpha\gamma}k_{\beta\delta} - k_{\alpha\delta}k_{\beta\gamma}) \;, \label{eq:gauss_eq} \\
{}^5\mathcal{R}_{ABCD} \mathcal{Z}^A_{,\alpha} \mathcal{Z}^B_{,\beta} \mathcal{Z}^C_{,\gamma} \eta^D &=& k_{\alpha[\beta;\gamma]} \;, \label{eq:codazzi}
\end{eqnarray}
where semicolons denote covariant derivatives with respect to the embedded metric $g_{\mu\nu}$. These equations serve as integrability conditions ensuring consistency between the bulk and the embedded geometry.

To ensure that matter remains confined to the embedded manifold, the energy-momentum tensor components of the bulk $T^{*}_{AB}$ must satisfy:
\begin{eqnarray}
\kappa^2_5 T^{*}_{\mu\nu} &=& 8\pi G T_{\mu\nu} \;, \\
\kappa^2_5 T^{*}_{\mu a} &=& 0 \;, \\
\kappa^2_5 T^{*}_{ab} &=& 0 \;, \label{eq:confined}
\end{eqnarray}
where $T_{\mu\nu}$ is the four-dimensional energy-momentum tensor and $G$ is Newton's gravitational constant. This ensures that only gravity probes the bulk while matter is effectively restricted to the embedded space imposed by Nash formula in Eq.\eqref{eq:nashdeformation}.

By varying the action in Eq.\eqref{eq:action} with respect to the bulk metric $\mathcal{G}_{AB}$, we obtain the bulk Einstein equations~\cite{MAIA20029,GDE, Maia_2007, gde2}
\begin{equation}\label{eq:EEbulk}
{}^5\mathcal{R}_{AB} - \frac{1}{2} \mathcal{G}_{AB} {}^5\mathcal{R} = \alpha^\star \mathcal{T}_{AB}\;,
\end{equation}
where $\alpha^\star$ is the bulk coupling constant and $\mathcal{T}_{AB}$ is the bulk energy-momentum tensor. For a flat or constant-curvature bulk, the associated Riemann tensor takes the form:
$$
{}^5\mathcal{R}_{ABCD} = K_\star \left( \mathcal{G}_{AC} \mathcal{G}_{BD} - \mathcal{G}_{AD} \mathcal{G}_{BC} \right) \;,
$$
with curvature scale $K_\star$ determining the sign of the cosmological constant in the bulk.

Assuming a flat bulk geometry, one obtains the following non-perturbed gravitational field equations from five dimensional bulk projected onto the embedded space-time that are simply written as
\begin{eqnarray}
G_{\mu\nu} - Q_{\mu\nu} &=& 8\pi G T_{\mu\nu} \;, \label{eq:embedded} \\
k_{\mu[\nu;\rho]} &=& 0 \;, \label{eq:embedded_codazzi}
\end{eqnarray}
where $Q_{\mu\nu}$ is a geometrically induced deformation tensor given by:
\begin{equation}\label{eq:Qmunu}
Q_{\mu\nu} = k^{\rho}_{\mu} k_{\rho\nu} - k_{\mu\nu} h - \frac{1}{2}(K^2 - h^2) g_{\mu\nu} \;.
\end{equation}
Here, $h = g^{\mu\nu} k_{\mu\nu}$ is the mean curvature, and $K^2 = k^{\mu\nu}k_{\mu\nu}$ represents the Gaussian curvature. More importantly, this tensor satisfies the conservation law
\begin{equation}\label{eq:Qconserved}
Q^{\mu\nu}{}_{;\mu} = 0 \;,
\end{equation}
assuring its dynamical consistency with the embedded geometry.

\section{A Gürses--Gürsey Metric Embedded in Five Dimensions}

In this work, we explore how extrinsic effects may manifest in a astrophysical gravitational system by considering a vacuum solution of a four-dimensional axially symmetric gravitational field. Using Boyer–Lindquist coordinates, we begin with a generalized Kerr metric, specifically the Gürses--Gürsey metric~\cite{gursey}, given by the line element:

\begin{eqnarray}\label{eq:gurseylement}
ds^2 & = \left( \frac{2 \, r f(r)}{\rho^2} - 1 \right) dt^2  -\frac{4 \, a r f(r) \sin^2\theta}{\rho^2} dt d\phi\nonumber\\ 
 &\;\;\;+ \frac{\rho^2}{\Delta}dr^2 + \rho^2 d\theta^2 + \frac{\Sigma}{\rho^2} \sin^2\theta d\phi^2\;,
\end{eqnarray}
where we define \( \rho^2 = a^{2} \cos^2\theta + r^{2} \), \( \Delta = a^{2} + r^{2} - 2 \, r f(r) \), and \( \Sigma = (r^2+a^2)^2 - a^2\Delta \sin^2\theta \). Here, in natural units, we represent as dimensionless parameters the spin parameter \( a = J/M \), with \( J \) as the angular momentum and \( M \) as the mass. The function \( f(r) \) is arbitrary and, in principle, encodes extrinsic geometric effects. Setting \( f(r) = M = \text{constant} \) recovers the standard Kerr solution, and further fixing \( a = 0 \) leads to the Schwarzschild metric.

To analyze how the extrinsic effects may influence the physics, we need to solve the field equations, namely Eq.\eqref{eq:embedded} and Eq.\eqref{eq:embedded_codazzi}. By Nash relation in Eq.\eqref{eq:nashdeformation}, small perturbations of an embedded manifold correspond to changes in the extrinsic curvature. Then, we propose the solution for $k_{\mu\nu}$ in a form 
\begin{equation} \label{eq:kmunuansatz}
k_{\mu\nu} = p_{\mu} g_{\mu\nu}, \quad \text{(no sum over \( \mu \))},
\end{equation}
where each $p_\mu=p_\mu (r,\theta,\phi,t)$ is a scalar function. The ansatz of Eq.\eqref{eq:kmunuansatz} implies that the extrinsic curvature is conformally related to the metric in each coordinate direction independently, due to the no sum over index $\mu$, implying that the deformations are proportional to the intrinsic geometry itself, meaning that the geometry of the embedded four--dimensional spacetime drives its own embedding curvature. Actually, this is a manifestation of Gauss and Codazzi equations of Eq.\eqref{eq:gauss_eq} and Eq.\eqref{eq:codazzi}, respectively. In other words, the dependence of $p_\mu$ as a function of coordinates $(r,\theta,\phi,t)$, introduces spacetime-dependent extrinsic curvature, suggesting local deformation of the embedded four--dimensional spacetime. Physically, this suggests the extrinsic geometry respects the symmetry of the intrinsic metric, allowing to study astrophysical consequences, e.g., for BH physics, star formation, and gravitational waves that all occur within locally curved spacetimes. A proper extension to cosmological symmetry at large-scale curvature of the universe is possible as well~\cite{MAIA20029,GDE,Maia_2007,gde2,Capistrano2021,capistrano2022,Capistrano2024}. Now, we are able to determine Eq.\eqref{eq:Qmunu} in a form
\begin{equation}
Q_{\mu\nu} = U_{\mu} g_{\mu\nu}, \quad \text{(no sum over \( \mu \))},
\end{equation}
where \( U_{\mu} \) is given by (see Ref.\cite{Capistrano2024b})
\begin{equation}
U_{\mu}=p^2_{\mu}-\left( \sum_{\alpha}p_{\alpha}\right)p_{\mu}
-\frac{1}{2}\left(\sum_{\alpha}p^2_{\alpha}-\left(\sum_{\alpha}p_{\alpha}\right)^2\right)\delta^\mu_\mu\;.    
\end{equation}
By imposing conservation \( \sum_{\nu} g^{\mu\nu}U_{\mu;\nu} = 0 \), we obtain two independent equations that result in the relations \( p_1 = p_2 = \alpha(r,\theta) \) and \( p_3 = p_4 = \beta(r,\theta) \). Assuming the isotropic case with \( \alpha = \beta \), we find
\begin{equation}
Q_{\mu\nu} = 3\alpha^2 g_{\mu\nu}.
\end{equation}
Substituting this into the field equation, we derive the radial equation for \( f(r) \):
\begin{equation}\label{eq:rad}
r \frac{d^2 f}{dr^2} + \frac{1}{2} \frac{df}{dr} + 6 \alpha^2 \rho^2 = 0.
\end{equation}
For \( \alpha = \frac{\sqrt{\alpha_0}}{\rho} \), the solution is
\begin{equation}\label{eq:fr}
f(r) = c_0 - 4\frac{\alpha_0}{r} + 2\sqrt{r}c_1,
\end{equation}
where the constant \( \alpha_0 \) reflects the extrinsic curvature effects, and \((c_0, c_1) \) are integration constants. To generalize the Kerr solution consistently, we need to ensure that Eq.\eqref{eq:gurseylement} remains physically viable, namely it should preserve asymptotic flatness and Lorentzian signature. This is guaranteed when we set $c_1 = 0$, since it eliminates linearly growing terms in $f(r)$ that would otherwise compromise the large-$r$ behavior. Hence, we simply have
\begin{equation}\label{eq:fr2}
f(r) = M - 4\frac{\alpha_0}{r}.
\end{equation}
In addition, we point out that this isotropy condition enforces isotropy in the embedding curvature, implying ( $k_{\theta\theta} = k_{\phi\phi}/\sin^2\theta )$. If we relax this condition, it introduces an anisotropy parameter $( \delta = (\alpha - \beta)/\alpha )$, leading to the potential
\begin{equation}\label{eq:anisotrparameter}
   \Psi(r, \theta) = 1 - \frac{2M}{r} + \alpha r^2 (1 + \delta \sin^2\theta).
\end{equation}
For $|\delta| \lesssim 10^{-3}$, the deformation of the shadow contour may become elliptical, yielding a shift in the shadow centroid of order ( $\Delta r/r \sim \delta/2$ ), and may be constrained by future interferometric observations.
 
\section{Equations of motion}
The Hamilton-Jacobi equation for a particle moving in this spacetime is:
\begin{equation}\label{eq:HJ01}
g^{\mu\nu} \frac{\partial S}{\partial x^\mu} \frac{\partial S}{\partial x^\nu} = -\mu^2\;,
\end{equation}
where \(\mu^2 \) is the mass-squared of the particle (\(\mu^2 = 0\) for photons and \(\mu^2 > 0\) for massive particles). The term $S$ is the action and \( g^{\mu\nu} \) is the inverse metric tensor. The, we start by assuming the separability of the action in the form
\begin{equation}\label{eq:HJ02}
S = \frac{1}{2} \mu^2 \lambda - E t + L \phi + S_r(r) + S_\theta(\theta)\;,
\end{equation}
where \( E = -p_t \) is the conserved energy of the particle, \( L = p_\phi \) is the conserved angular momentum about the symmetry axis. The generalized momentum is given by \(p^\mu = g^{\mu\nu} \partial_\nu S\). From the metric structure and standard algebra, we use Carter's separation method. From Eq.\eqref{eq:HJ02}, we have
\begin{eqnarray}
p_t &&= \frac{\partial S}{\partial t} = -E = g_{tt} \dot{t} + g_{t\phi} \dot{\phi}\;,\\
p_\phi &&= \frac{\partial S}{\partial \phi} = L = g_{\phi t} \dot{t} + g_{\phi\phi} \dot{\phi}\;.
\end{eqnarray}
These give a linear system:
\begin{eqnarray}
-E &&= g_{tt} \dot{t} + g_{t\phi} \dot{\phi}\;,\\
L &&= g_{t\phi} \dot{t} + g_{\phi\phi} \dot{\phi}\;.
\end{eqnarray}
Solving this system for $\dot{t}$ and $\dot{\phi}$, one finds
\begin{eqnarray}
\dot{t}&& = \frac{\rho^2 (2aLrf - E \Sigma)}{(2rf - \rho^2)\Sigma - 4a^2r^2 f^2 \sin^2\theta}\;,\\
\dot{\phi}&& = \frac{\rho^2 (L \Sigma - 2a E r f)}{(2rf - \rho^2)\Sigma - 4a^2 r^2 f^2 \sin^2\theta}\;.
\end{eqnarray}
In the equatorial plane $\theta = \pi/2$, and assuming a test particle ($\mu = 1$), we have
\begin{eqnarray}
\dot{t} &&= \frac{E(r^2 + a^2) + 2a f(r)(aE - L)/r}{r^2 + a^2 - 2rf(r)}\;,\\
\dot{\phi}&& = \frac{L + 2 f(r)(aE - L)/r}{r^2 + a^2 - 2rf(r)}\;.
\end{eqnarray}
From separation of Eq.\eqref{eq:HJ02} and assuming $\mu = 1$, the radial part becomes
\begin{equation}
\rho^4 \dot{r}^2 = [ (r^2 + a^2) E - aL ]^2 - \Delta [ r^2 + \mathcal{K}]\;,
\end{equation}
where $\mathcal{K}$ is the Carter constant $\mathcal{K} = Q+ (aE - L)^2$.  The term $Q$ is the part that depends only on $\theta$ and it comes from the angular separation of the Hamilton-Jacobi equation. The term $(aE - L)^2$ appears naturally in both radial and angular parts due to separability. For the angular motion, we obtain
\begin{equation}
\rho^4 \dot{\theta}^2 =\mathcal{K} - a^2 \cos^2\theta (E^2 - 1) - L^2 \cot^2\theta\;.
\end{equation}
This is the standard result from separating the angular part of the Hamilton-Jacobi equation in Kerr spacetime, generalized to $f(r)$ via $\rho^2$. For the case $\alpha_0\rightarrow 0$ that makes $f(r) = m$, we obtain the correct reduction to Kerr $a\neq 0$ and Schwarzschild $a = 0$ space-times, respectively.

\subsection{Effective potentials and horizons}
The effective potential for radial motion is derived from the radial equation \( R(r)=\rho^4 \dot{r}^2 \). The radial equation for geodesics in the given metric is:
\begin{equation}
R(r) = \left( (r^2 + a^2) E - a L \right)^2 - \Delta \left( Q + r^2 \right).
\end{equation}
In a similar fashion, we write the angular equation with $\rho^4 \dot{\theta}^2 = \Theta(\theta)$, and obtain
\begin{equation}
\Theta(\theta) = \mathcal{K} - \left[ \frac{L^2}{\sin^2\theta} + a^2 (E^2 - 1) \cos^2\theta \right]\;.
\end{equation}
For a particle to move radially, we impose the condition \(R(r) \geq 0\). Then, the radial geodesic equation is written as
\begin{equation}
\left( \frac{dr}{d\lambda} \right)^2 + V_{\text{eff}}(r) = 0,
\end{equation}
where \( \lambda \) is an affine parameter. Thus, the effective potential \( V_{\text{eff}} \) is derived from \( R(r) \)
\begin{equation}
V_{\text{eff}}(r) = - \frac{R(r)}{\rho^4} .
\end{equation}
For simplicity, we consider equatorial motion (\( \theta = \pi/2, Q=0 \)). Substituting \( R(r) \) and simplifying, we get
\begin{equation}
V_{\text{eff}}(r) = \frac{\Delta \left(r^2+ (L - a E)^2  \right) - \left( (r^2 + a^2) E - a L \right)^2}{r^4}.
\end{equation}
This can be rewritten in a Kerr-like form as
\begin{equation}
V_{\text{eff}}(r) = \frac{ \Delta \left( r^2 + (L - a E)^2 \right) - \left( (r^2 + a^2) E - a L \right)^2 }{r^4},
\end{equation}
where \( \Delta = r^2 + a^2 - 2Mr + 8\alpha_0 \). The Schwarzschild limit is obtained with (\( a = 0, \alpha_0 = 0 \)) given the known expression  
\begin{equation}
  V_{\text{eff}}(r) = \left(1 - \frac{2M}{r}\right) \left(1 + \frac{L^2}{r^2}\right) - E^2\;,
\end{equation}
and for (\( \alpha_0 = 0 \)), we obtain the standard Kerr BH. In Table (\ref{tab:horizons}), we present the inner and outer horizons \(r_{\pm}\) for a BH with mass \(M = 1\) for a fixed \(\alpha_0 = 0.0069\). In the subsection \eqref{sub:cosmolink}, we discuss such adopted value for \(\alpha_0\) and its link to a cosmological scale.  

The horizons in this spacetime can be found by applying the standard procedure with
\(\Delta = a^2 + r^2 - 2r f(r) = 0\) and one has
\begin{equation}\label{eq:horizons}
r_{\pm} = M \pm \sqrt{M^2 - (a^2 + 8\alpha_0)}.
\end{equation}
These are the event horizon (\( r_{+} \)) and inner horizon (\( r_{-} \)), respectively. If \( M^2 > a^2 + 8\alpha_0 \), two horizons exist. If \( M^2 = a^2 + 8\alpha_0 \), there is one degenerate horizon (extremal case). If \(M^2 < a^2 + 8\alpha_0 \), there is no horizon (naked singularity). For \( \alpha_0 = 0 \), we recover the standard Kerr horizons , while \( \alpha_0 \) resembles a tidal ``extrinsic charge'' akin to the Kerr–Newman solution. 

\subsection{Parameter inheritance: matching local and cosmological scales}\label{sub:cosmolink}
A cosmological model was developed in \cite{Capistrano2024} using the main field equations of Eq.\eqref{eq:embedded} and \eqref{eq:embedded_codazzi} under cosmic perturbations. In the conformal Newtonian gauge, the Friedmann–Lemaître–Robertson–Walker~(FLRW) metric is given by
\begin{equation}\label{eq:scalarpertmetric2}
ds^2 = a^2 [-(1+ 2\Psi) d\tau^2 + (1-2\Phi) dx^i dx_i] \;,
\end{equation}
where $a=a(t)$ is the scale factor. The functions $\Psi=\Psi(\vec{x},\tau)$ and $\Phi=\Phi(\vec{x},\tau)$ denote the Newtonian potential and the Newtonian curvature in conformal time $\tau$ that is defined as $d\tau=dt/a(t)$. As a result, the modified Friedman equation from is written as
\begin{equation}\label{eq:Friedman2}
H^2=\frac{8}{3}\pi G \left(\bar{\rho}_{m}+\bar{\rho}_{ext}\right)\;\;,
\end{equation}
where $\bar{\rho}_{m}$ is the matter energy density and $\bar{\rho}_{ext}(a)$ is the extrinsic energy density that is given by
\begin{equation}\label{eq:extdensitya1}
\bar{\rho}_{ext}(a)=\bar{\rho}_{ext}(0)a^{2\beta_0-4}\;,
\end{equation}
with $\bar{\rho}_{ext}(0)=\frac{3}{8\pi G} b_0^2$. The $\beta_0$ parameter is an integration constant and $b_0$ comes from the extrinsic influence that mediates the contribution of extrinsic curvature. Thus, we write the relevant the gauge invariant perturbed field equation in the Fourier \emph{k}-space wave modes as
\begin{eqnarray}
&&k^2\Phi_k + 3 \mathcal{H} \left(\Phi^{'}_k + \Psi_k \mathcal{H} \right)= -4\pi G a^2 \delta \rho_k + \chi(a)\Psi_k , \label{tensorcompo00kspace}\\
&&\Phi^{'}_{k}+ \mathcal{H}\Psi_k= -4\pi G a^2(\bar{\rho}+P) \frac{\theta}{k^2}\;,\label{tensorcomp0ikspace}\\
&&\mathcal{D}_k + \frac{k^2}{3}(\Phi_k-\Psi_k)= -\frac{4}{3}\pi G a^2 \delta \bar{P}-\frac{1}{2}a^2 \delta Q^{i}_{i}\;, \label{tensorcompijkspace}\\
&&k^2(\Phi_{k}-\Psi_k)= 12\pi G a^2(\bar{\rho}+P) \sigma,\label{offdiagonal}
\end{eqnarray}
where $\theta= ik^j\delta u_{\parallel j}$ denotes the divergence of fluid velocity in \emph{k}-space, and $\mathcal{D}_k$ denotes $\mathcal{D}_k=\Phi^{''}_{k} + \mathcal{H}(2\Phi_k+\Psi_k)' + (\mathcal{H}^2+2\mathcal{H}')\Psi_k$. The function $\chi(a)$ is expressed in terms of the cosmological parameters and reads \begin{equation}\label{eq:dimenHub}
\chi(a)= \frac{9}{2}\gamma_s \frac{H_0^2}{\Omega_{rad}(a)}\Omega_{rad}(0) \Omega_{ext}(a) \;,
\end{equation}
where $\gamma_s$ is a dimensionless coupling related to $b_0$ that controls the strength of extrinsic curvature perturbations.

The use of the set of perturbed equations in Eqs.(\ref{tensorcompo00kspace}), (\ref{tensorcomp0ikspace}, (\ref{tensorcompijkspace}) and (\ref{offdiagonal}) is simplified after some algebra into the following set of equations 
\begin{eqnarray}
&&k^2\Psi_k = -4\pi G a^2 \mu(a,k)\rho \Delta\;,\label{muequation1}\\
&&k^2(\Phi_{k}+\Psi_k)= -8\pi G a^2 \Sigma(a,k)\rho\Delta\;,\label{sigmaequation1}
\end{eqnarray}
where $\rho \Delta= \bar{\rho}\delta+ 3\frac{\mathcal{H}}{k}(\bar{\rho}+P)\theta$. The set of Eqs.(\ref{muequation1}) and (\ref{sigmaequation1}) are valid for all times. When anisotropic stress is neglected, $\mu(a,k)$ and $\Sigma(a,k)$ functions can be written as
\begin{eqnarray}
&&\mu(a,k) = \frac{1}{1-\frac{\chi(a)}{k^2}} \;,\label{muequation}\\
&&\Sigma(a,k)= \frac{1}{2}\left[1+\mu(a,k)\left(1+\frac{\chi(a)}{k^2}\right)\right]\;.\label{sigmaequation}
\end{eqnarray}

Using the definition of the slip function $\gamma(a,k)=\frac{\Phi}{\Psi}$, from Eqs. (\ref{muequation}) and (\ref{sigmaequation}), one easily obtains
\begin{equation}\label{eq:mg3functions}
\Sigma(a,k)= \frac{1}{2}\mu(a,k)(1+\gamma(a,k))\;.
\end{equation}
When the extrinsic term $\gamma_s\rightarrow 0$ in order to recover GR correspondence, one obtains the standard GR limit as  $\Sigma(a,k)=\mu(a,k)$ and $\gamma(a,k)=1$. To obey solar constraints, $\gamma_s$ in $\mu(a,k)$ function must comply with the condition 
\begin{equation}\label{eq:gammas2}
\gamma_s< \frac{0.222 k^2_p}{H_0^2\Omega_{ext(0)}}\;,
\end{equation}
at pivot scale wave-number $k_p$.  

Using the latest NPIPE Planck DR4 likelihoods~\cite{Carron_2022,Rosenberg:2022sdy}, the BICEP2/Keck collaboration \cite{BICEPKeck} and a junction of Large-scale structure (LSS) catalogue with 6dF Galaxy Survey~\cite{6dFGalaxy}, the Seventh Data Release of SDSS Main Galaxy Sample (SDSS DR7 MGS)~\cite{Ross:2014qpa} and clustering measurements of the Extended Baryon Oscillation Spectroscopic Survey (eBOSS) associated with the SDSS's Sixteenth Data Release~\citep{Alam:2016hwk} at pivot scale $k_p=0.05$Mpc$^{-1}$ with $68\%$ confidence level, we obtained $10^{7}\gamma_s= 0.03\pm 0.27$~\cite{Capistrano2024}. This joint likelihood analysis was computed with an
\texttt{MGCAMB-II}~\cite{mgcamb2023} with \texttt{Cobaya}~\cite{cobaya} sampler to compute MCMC runs for a Gelman-Rubin convergence criteria \(R=0.02\) employing the $\mu-\Sigma$ parametrization given by
\begin{eqnarray}
&&\mu(a,k)= 1 + \mu_0 \frac{\Omega_{DE}}{\Omega_{DE(0)}}\;,\label{eq:mucamb1}\\
&& \Sigma(a,k)= 1 + \Sigma_0 \frac{\Omega_{DE}}{\Omega_{DE(0)}} \;.\label{eq:sigmacamb1}
\end{eqnarray}
Here, $\Omega_{DE}(a)$ is denoted as the extrinsic contribution $\Omega_{ext}(a)$ which means that $\Omega_{DE(0)}=\Omega_{ext}(0)=1-\Omega_{m(0)}$. The form of Eqs. (\ref{eq:mucamb1}) and (\ref{eq:sigmacamb1}) in this parametrization is obtained from expanding the function $\chi(a)<<1$ in the denominator of Eq. (\ref{muequation}). The priors on the baseline parameters utilized in our analysis are detailed in Table \ref{tab:priors}. Then, we calculate \(\alpha_0\) as a derived parameter with a phenomenological scaling relation \begin{equation}\label{eq:extomega}
\alpha_0 = 10^{7}\gamma_s \Omega_m\, h,
\end{equation}
where \( h \) is the dimensionless Hubble factor. From the form of Eq.\eqref{eq:extomega}, we guarantee that the extrinsic \(\alpha_0=0.0069\pm 0.0621\) parameter inherits a cosmological origin by means of $\gamma_s$ parameter.
\begin{table}
\centering
\begin{tabular}{l|l}
\hline \hline
Parameter                                   & Prior        \\ \hline
$ \Omega_b h^2$                             & $\mathcal{U}[0.017,0.027]$ \\
$ \Omega_c h^2$                             & $\mathcal{U}[0.09,0.15]$ \\
$\theta_\mathrm{MC}$                        & $\mathcal{U}[0.0103, 0.0105]$       \\
$\tau_{\rm reio}$                           & $\mathcal{N}[0.065,0.0015]$   \\
$\log(10^{10} A_\mathrm{s})$                & $\mathcal{U}[2.6,3.5]$       \\
$n_{s}$                                     & $\mathcal{U}[0.9, 1.1]$     \\
$10^{8}$ $\gamma_{s}$       & $\mathcal{U}[-1, 1]$     \\
\hline
\end{tabular}
\caption{The cosmological parameters along with their respective priors employed in the parameter estimation analysis. }
\label{tab:priors}
\end{table}
Although $\alpha_0$ was derived using Planck NPIPE and BICEP2/Keck data, switching to DESI~\cite{DESI2024VI} with $\Omega_m= 0.33$, $h = 0.68$, it changes the value of $\alpha_0$ parameter by only $4\%$, showing good stability confirming that cosmological uncertainties have negligible effect on observable scales.

As we have a value for \(\alpha_0\), it leads to a peculiar behavior of the horizons determined by Eq.\eqref{eq:horizons}. As \(|a|\) increases, the horizons move closer together. When \(|a| = \sqrt{0.952} \approx 0.997\), the horizons coincide (\(r_{+} = r_{-} = 1\)), corresponding to an extremal black hole. For \(|a| > 0.997\), the square root becomes imaginary, and no horizons exist (naked singularity). The term \(8\alpha_0\) reduces the value under the square root, effectively decreasing the allowed range of \(a\) for which horizons exist. This suggests that \(\alpha_0\) turns the formation of horizons more restrictive compared to the standard Kerr case (where \(\alpha_0 = 0\) and the extremal limit is \(|a| \leq 1\)). The results are summarized in table The extremal limit occurs at \(|a| \approx 0.997\), beyond which no horizons exist. Then,  $\alpha_0$ is small enough to allow astrophysically realistic BHs. Hence, the presence of $\alpha_0>0$ slightly reduces the extremal spin compared to Kerr, making horizon formation more restrictive. 

\begin{table*}[ht!]
\centering
\caption{Inner and outer horizons \(r_{\pm}\) for a BH with mass \(M = 1\) and the parameter \(\alpha_0 = 0.0069\). }
\label{tab:horizons}
\begin{tabular}{cccc}
\toprule
Spin \(a\) & Outer horizon \(r_{+}\) & Inner horizon \(r_{-}\) & Horizon condition \\
\midrule
\(0.0\)    & \(1.976\)              & \(0.024\)              & Distinct horizons \\
\(0.5\)    & \(1.838\)              & \(0.162\)              & Distinct horizons \\
\(0.8\)    & \(1.492\)              & \(0.508\)              & Distinct horizons \\
\(0.976\)  & \(1.000\)              & \(1.000\)              & Extremal (merged) \\
\(1.0\)    & ---                    & ---                    & No horizons (naked singularity) \\
\bottomrule
\end{tabular}
\end{table*}

\begin{figure*}[t!]
\centering
\includegraphics[width=6in, height=2.6in]{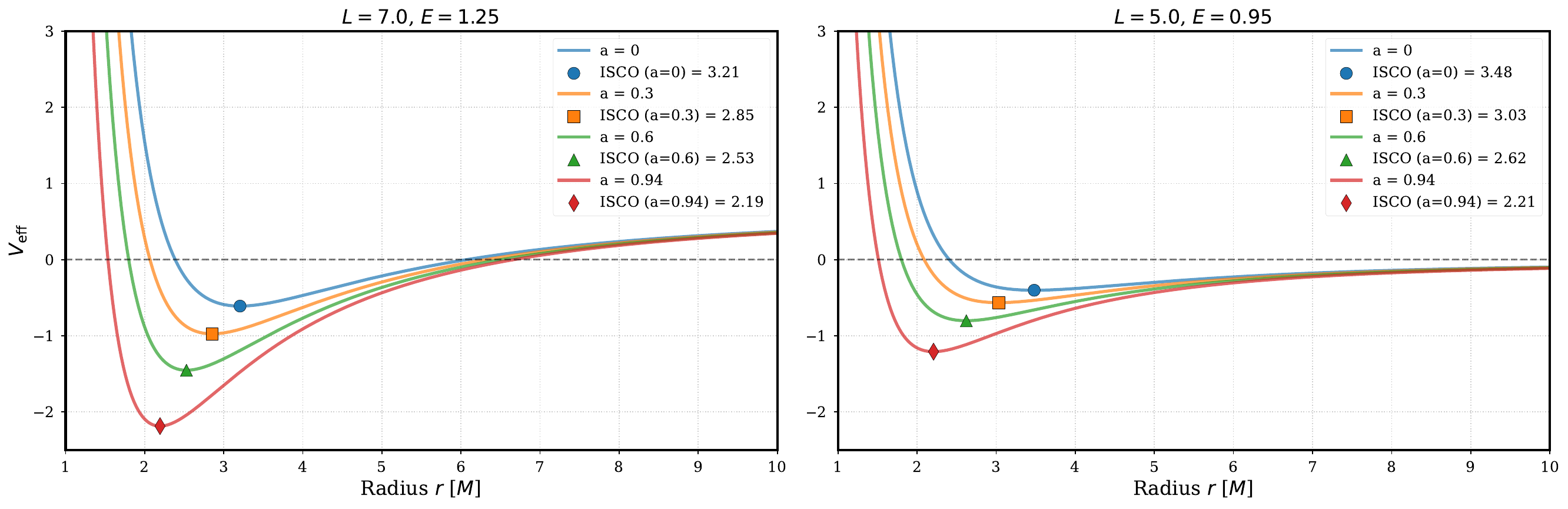}
\caption{Effective potential for different values of $L$ and $E$ for massive particles $\mu=1.0$ for each spin $a$. The ISCO are also represented for each spin $a$, accordingly.} \label{fig:effpot1}
\end{figure*}

In Fig.\ref{fig:effpot1}, it is shown two plots for the effective potential $V_{\text{eff}}(r)$ as a function of radius $(r)$ that is plotted for various values of the BH spin parameter $a$, namely, $a = (0,\, 0.3,\, 0.6,\, 0.94)$ represented, in this order, by the curves and the indicators: red line with diamond, green line with triangle, orange line with square and blue line with a sphere. Each case also includes the ISCO radius for each spin. In the left and right panels, we have $L = 7.0, E = 1.25$ and  $L = 5.0, E = 0.95$, respectively. These values were tested in order to obtain stable circular orbits as shown by the minima of $V_{\text{eff}}$ in the panels.

The overall pattern follows that as $a$ increases, the ISCO radius decreases indicating prograde orbits pulled inward. In the left panel with $(L = 7.0, E = 1.25)$, the potential well is clearly deeper with higher $a$, indicating more tightly bound orbits and the ISCO moves inward. In the right panel with $(L = 5.0, E = 0.95)$, we have shallower potential wells, indicating less tightly bound orbits. The ISCOs are further out, especially for the case $a = 0$. These results reinforce the overall trends as expected for high-spin Kerr-like BHs: the inner edge of the disk (located at the ISCO) is closer to the event horizon, allowing more efficient energy extraction and hotter inner disk regions. The gravitational redshift and temperature of emitted radiation increase with decreasing ISCO. High-spin BHs show higher luminosity due to deeper gravitational wells. On the other hand, the decreasing ISCO with spin allows stable orbits in regions of stronger gravity, impacting stability criteria for particles and matter near the black hole. As expected, frame-dragging is more pronounced at high $a$, which is significant in mechanisms like the Blandford--Znajek process~\cite{BlandfordZnajek1977} for jet production. 

In the following, we propose to link local strong-field effects with cosmic-scale corrections, provided that certain consistency conditions are satisfied. The parameter $\alpha_0$ is then regarded as a renormalized constant, that is, a local remnant of cosmological geometry, constant in the BH vicinity but fixed by global (cosmic) conditions. Moreover, the parameter $\alpha_0$ is constant in time at late times and the modifications do not spoil the local asymptotic flatness of the spacetime, as provided by Eq.\eqref{eq:fr2}, modeling an isolated rotating object, but not directly embedded in FLRW. This ensures that the spacetime remains stationary and asymptotically flat, consistent with the properties of an isolated Kerr-like solution. In particular, $\alpha_0$ modifies only the extremal spin condition and does not introduce a radial dependence that would spoil local geometry. This is analogous to scenarios in which global cosmological parameters (e.g., a small cosmological constant) leave the local Schwarzschild or Kerr solutions approximately unchanged over astrophysical scales.

\subsection{Modified Komar mass}
For Kerr spacetime, the Komar mass at spatial infinity matches the ADM mass $m$. On the other hand, for our modified function given by Eq.\eqref{eq:fr2}, the asymptotic behavior of the $g_{tt}$ component of Eq.\eqref{eq:gurseylement} becomes
\begin{equation}
g_{tt} \approx -1 + \frac{2rm}{r^2} - \frac{8\alpha_0}{r^2} + \mathcal{O}(1/r^3)\;.
\end{equation}
This implies that the ADM mass is still $m$, but the Komar mass contains higher-order corrections from $\alpha_0$. To calculate the Komar mass at finite radius, we use
\begin{equation}\label{eq:komar}
M_K(r) = \frac{1}{4\pi} \int_S \left( n_\mu \sigma_\nu \nabla^\mu \xi^\nu \right) \sqrt{\sigma} \, d\theta d\phi,
\end{equation}
where $\xi^\nu = \delta^\nu_t$, $n_\mu$ is unit timelike normal to $\Sigma_t$, $\sigma_\nu$ is spacelike normal to the 2-sphere $S$. After simplifying, reducing to static, axisymmetric case, we get
\begin{equation}
M_K(r) = \frac{1}{2} r^2 \partial_r g_{tt}.
\end{equation}
Using the component $g_{tt}$ of Eq.\eqref{eq:gurseylement}, we obtain the Komar mass at radius $r$ as
\begin{equation}\label{eq:komarmass}
M_K(r) = m - \frac{8\alpha_0}{r}.
\end{equation}
This interpolates between $m$ (at infinity) and $M_K(r)\to \infty$ as r$\to 0$. Thus, the effect of $\alpha_0$ is repulsive, consistent with its interpretation as an effective ``tidal charge''. At infinity, $M_K(\infty) = m$ that is the standard ADM mass. Inasmuch as we have $r \to 0$, then $M_K(r) \to -\infty$ if $\alpha_0 > 0$. Hence, the correction reduces the Komar mass at small $r$, indicating an attractive contribution. In the traditional braneworld context, this behavior resembles the Dadhich–Maartens–Papadopoulos–Rezania (DMPR) solution~\cite{Dadhich2000}, the Schwarzschild-like metric has a correction
\begin{equation}
g_{tt} = -\left(1 - \frac{2M}{r} + \frac{q}{r^2}\right)\;,
\end{equation}
where the term $\frac{q}{r^2}$ arises from bulk effects projected onto the brane, interpreted as a tidal charge. Different from the standard Reissner–Nordström term which the charge $q$ naturally comes from the electromagnetic field contribution, the sign of $q$ in this context can also be negative, allowing for a stronger gravitational field than in standard GR solution.

\subsection{Effective energy-momentum tensor~(EMT)}
For an exercise of interpretive purposes, we pack the known geometric corrections as effective matter. Then, we define an effective energy-momentum tensor of Eq.\eqref{eq:embedded} via
\begin{equation}
    T^{\text{eff}}_{\mu\nu} = \frac{1}{8\pi G} G_{\mu\nu}.
\end{equation}
For simplicity, we adopt the nonrotation limit $a \to 0$. Thus, the metric in Eq.\eqref{eq:gurseylement} becomes
\begin{equation}
ds^2 = -\left(1 - \frac{2rf(r)}{r^2} \right) dt^2 + \left(1 - \frac{2rf(r)}{r^2} \right)^{-1} dr^2 + r^2 d\Omega^2.
\end{equation}
Using standard expressions for static, spherically symmetric metrics with Eq.\eqref{eq:fr2}, we find
\begin{equation}
g_{rr}^{-1} = 1 - \frac{2f(r)}{r} = 1 - \frac{2m}{r} + \frac{8\alpha_0}{r^2}.
\end{equation}
So after straightforward calculation, we obtain the components $G^t_t = -8\pi \rho_{\text{eff}} = -\frac{16\alpha_0}{r^4}$,
$G^r_r = 8\pi p_r^{\text{eff}} = -\frac{16\alpha_0}{r^4}$,
$G^\theta_\theta = 8\pi p_\theta^{\text{eff}} = \frac{16\alpha_0}{r^4}$, where the energy density and pressure of extrinsic contribution are denoted by $\rho_{\text{eff}}$ and $p_r^{\text{eff}}$, respectively. Therefore, we obtian the effective EMT that corresponds to anisotropic matter with
\begin{equation}
\rho = -p_r = \frac{2\alpha_0}{\pi r^4}, \quad p_\theta = -\rho.
\end{equation}
This also resembles the EMT of a tidal charge in DMPR solution, but now, they are derived from full embedding conditions. Then, we can also explore the physical energy conditions. The Weak Energy Condition (WEC) requires $\rho \geq 0$, $\rho + p_i \geq 0$, and we have
\begin{equation}
\rho = \frac{2\alpha_0}{\pi r^4} > 0,
\rho + p_r = 0,
\rho + p_\theta = 0,
\end{equation}
which is marginally satisfied. The Null Energy Condition (NEC) requires that $\rho + p_i \geq 0$, and in this case is also marginal. Moreover, the Strong Energy Condition (SEC) requires $\rho + \sum p_i \geq 0$
but we have
\begin{equation}
\rho + p_r + 2p_\theta = \rho - \rho + 2(-\rho) = -2\rho < 0\;,
\end{equation}
which is clearly violated. The Dominant Energy Condition (DEC) requires that $\rho \geq |p_i|$ $\rho = |p_r| = |p_\theta|$ which is marginally satisfied. The effective matter violates the SEC, implying that this geometry allows repulsive gravitational effects, often associated with dark energy. The marginal satisfaction of WEC, NEC, and DEC suggests the matter is non-standard, but still physically plausible in extended theories. The radial decay as $\sim 1/r^4$ points to a tidal-like or higher-dimensional origin. 

\section{Shadows and visibility}
\begin{figure*}[t!]
    \centering
    \textbf{\hspace{0.5cm} $i=17^\circ$ \hspace{3cm} $i=45^\circ$ \hspace{3.8cm} $i=85^\circ$} \\
     \rotatebox{90}{\textbf{$\vphantom{\beta(M)}$}}
    \includegraphics[width=2.1in, height=2.1in]{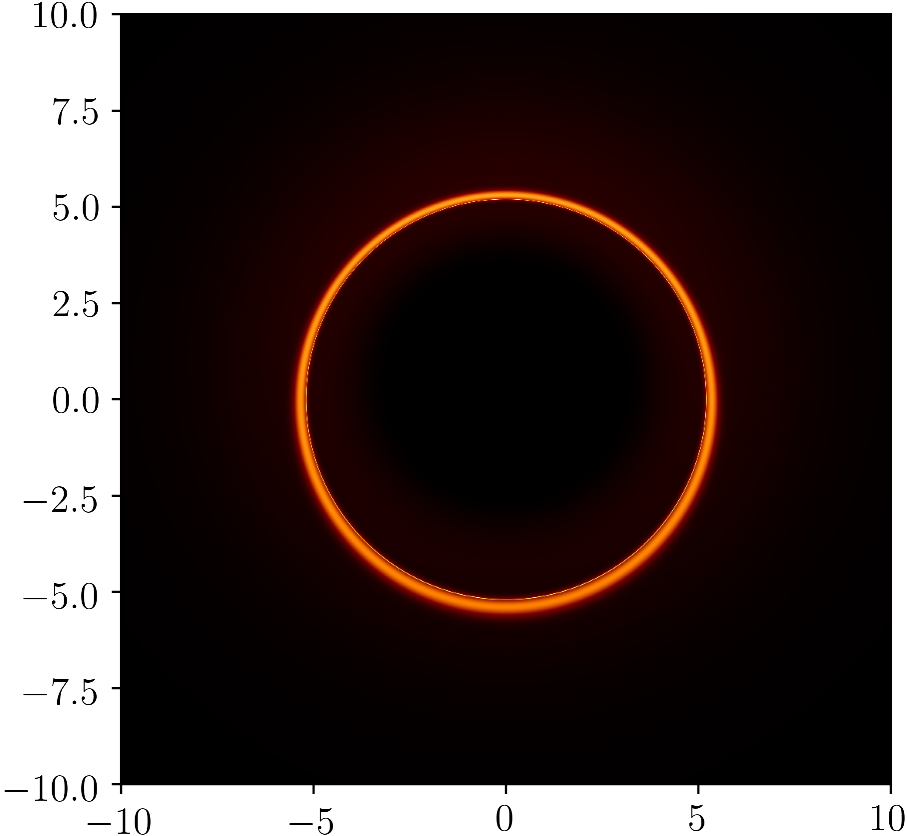}
    \includegraphics[width=2.1in, height=2.1in]{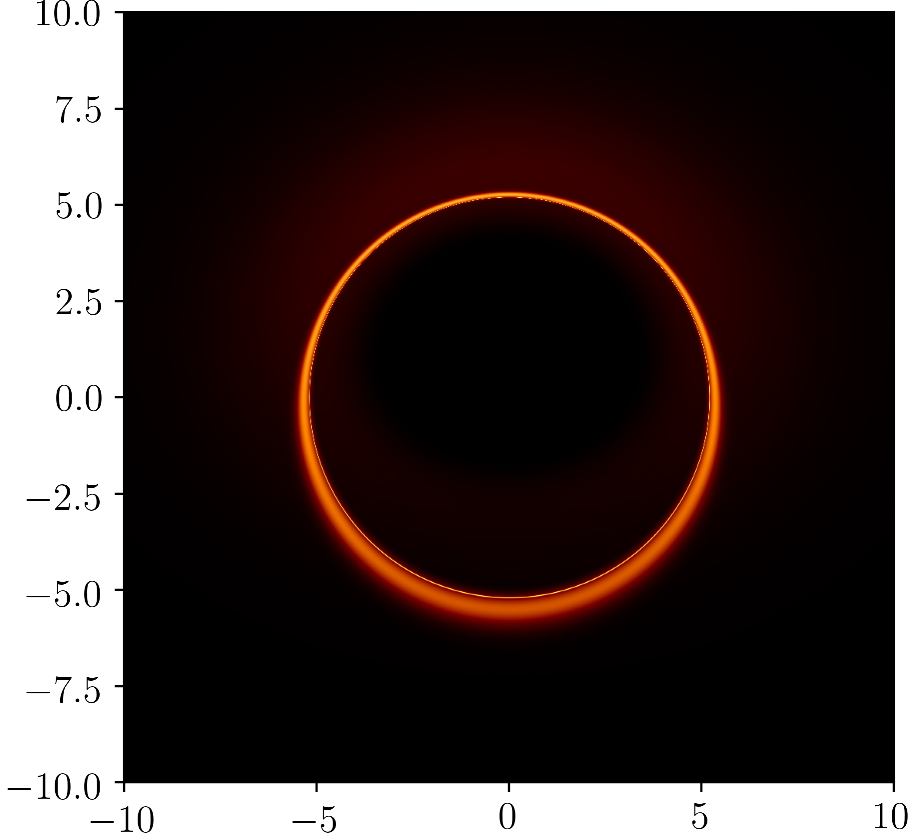}
    \includegraphics[width=2.1in, height=2.1in]{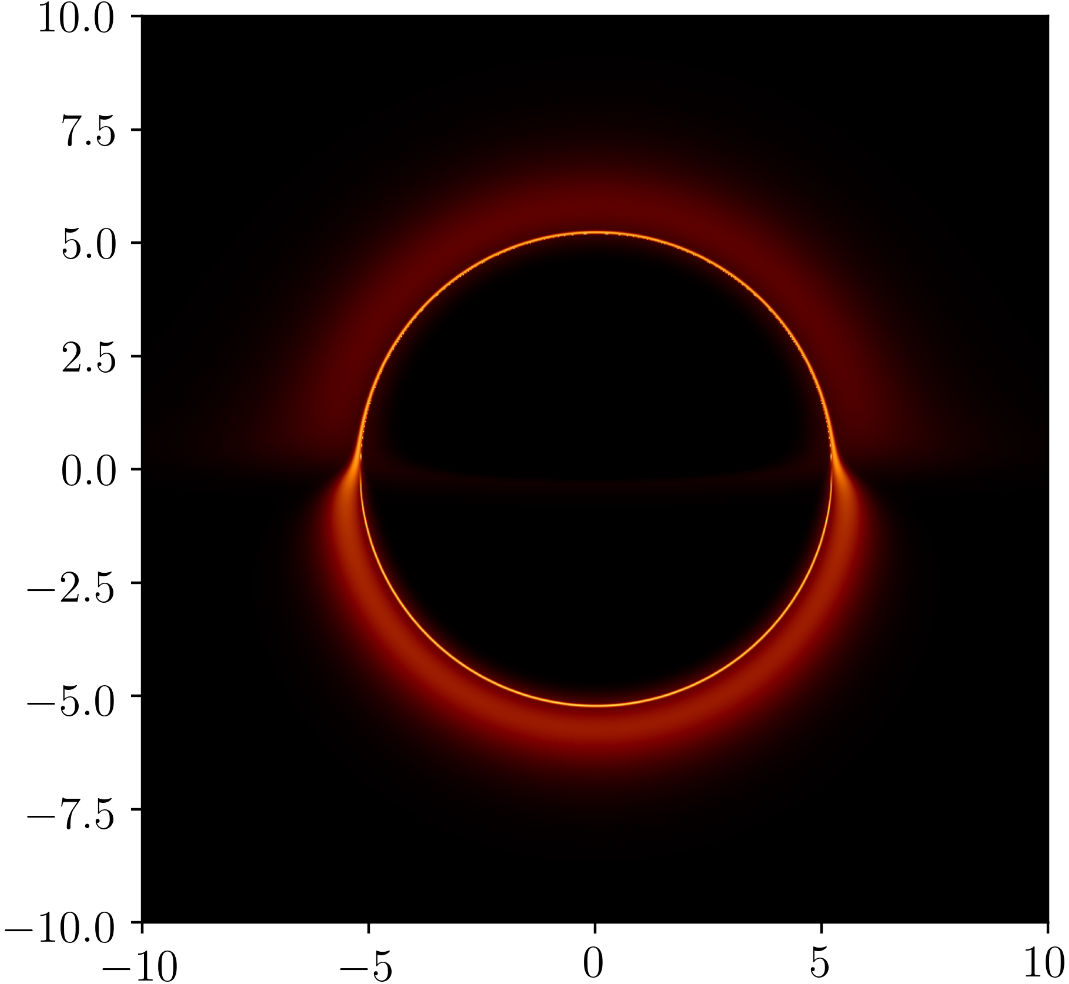} \\
    \rotatebox{90}{\textbf{ \large $\hspace{0.5cm}\;\;\;\;\;\beta (M)$}}
    \includegraphics[width=2.1in, height=2.1in]{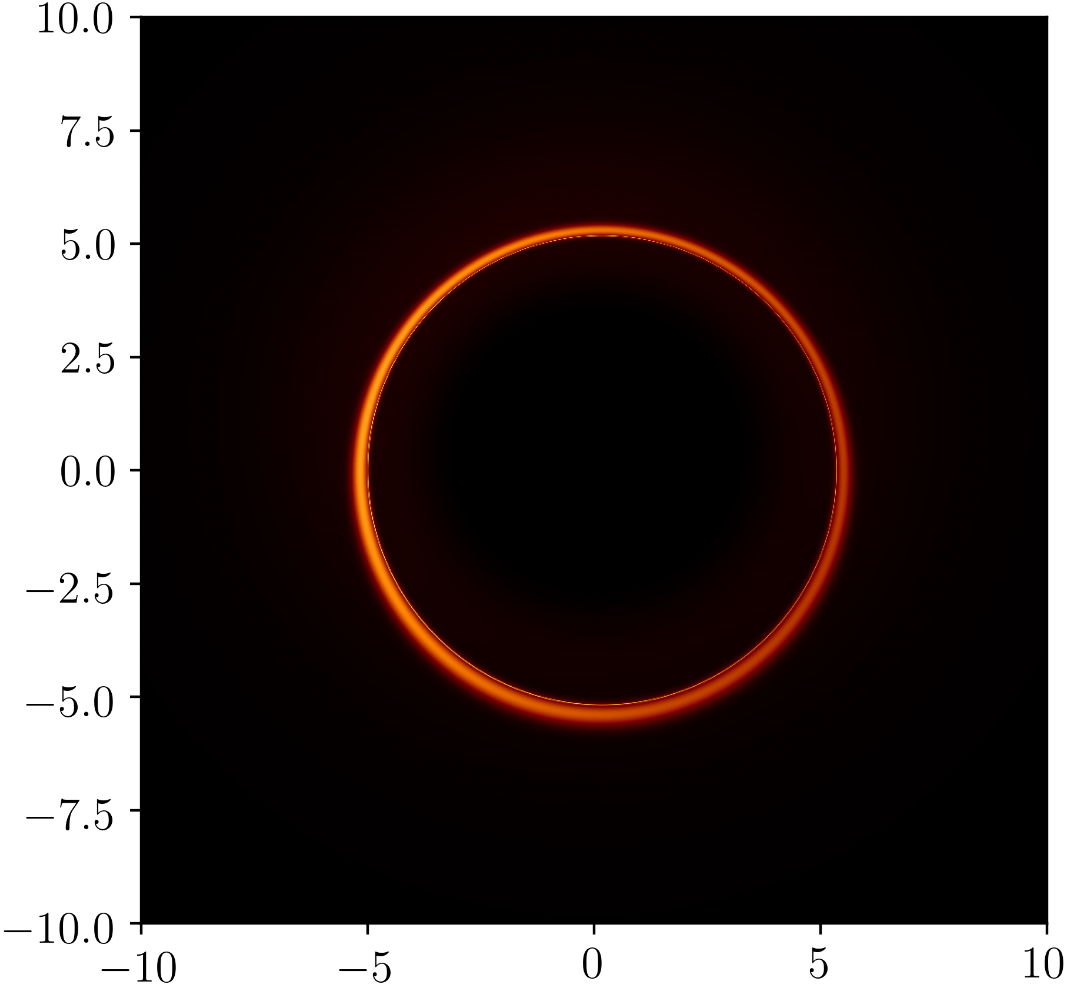}
    \includegraphics[width=2.1in, height=2.1in]{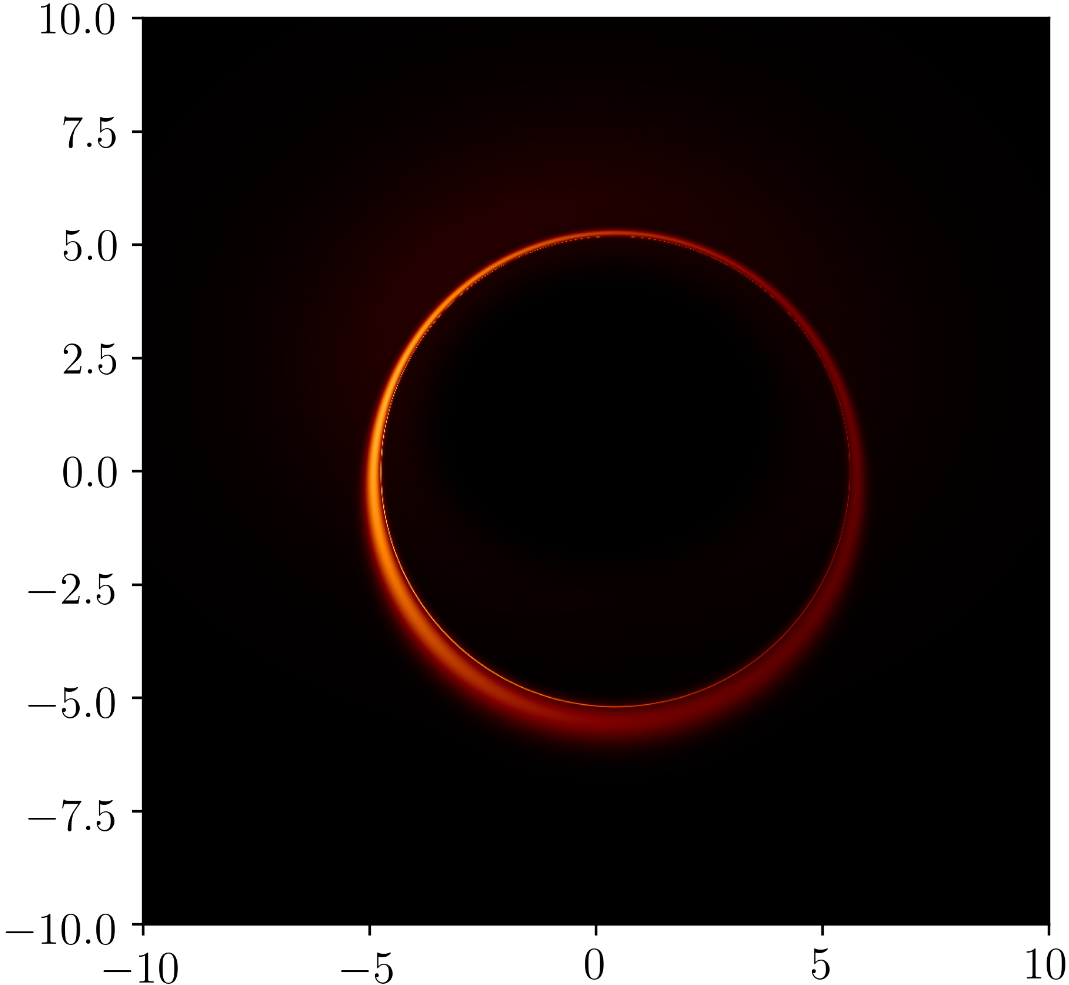}
    \includegraphics[width=2.1in, height=2.1in]{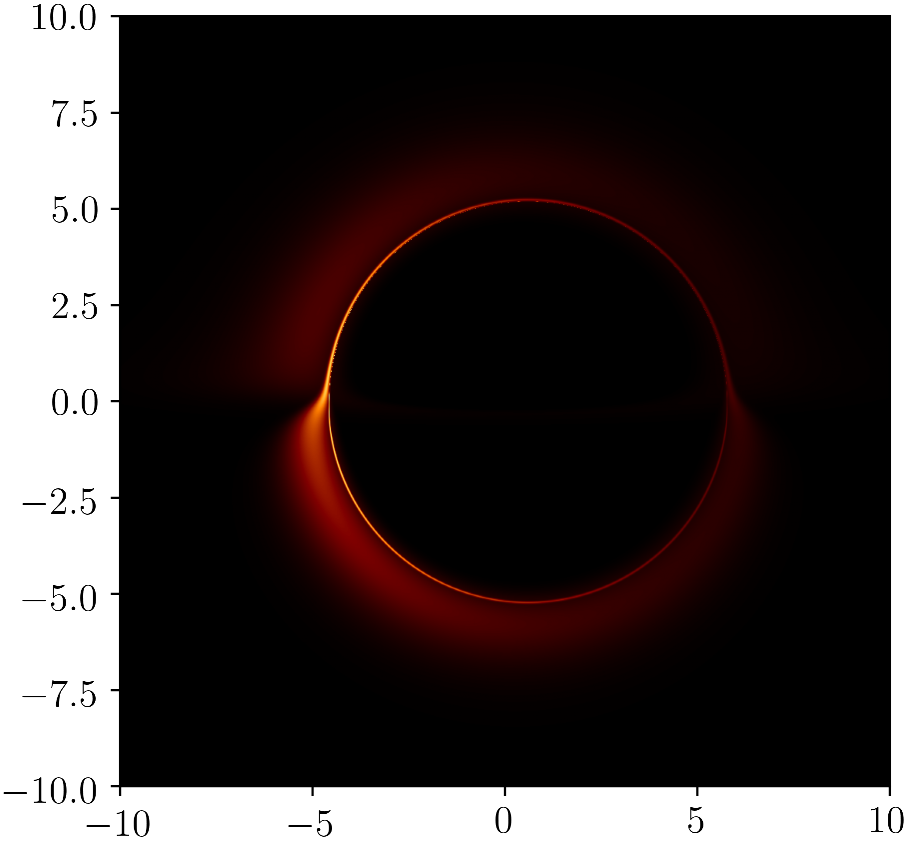} \\
    \rotatebox{90}{\textbf{$\vphantom{\beta(M)}$}}
    \includegraphics[width=2.1in, height=2.1in]{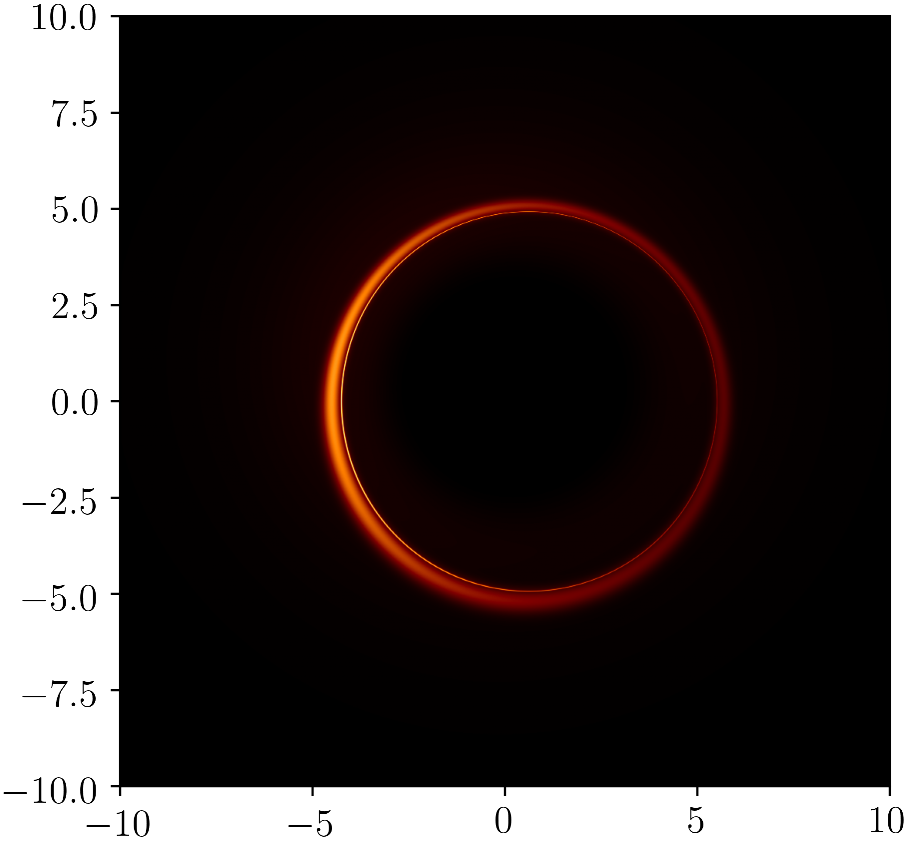}
    \includegraphics[width=2.1in, height=2.1in]{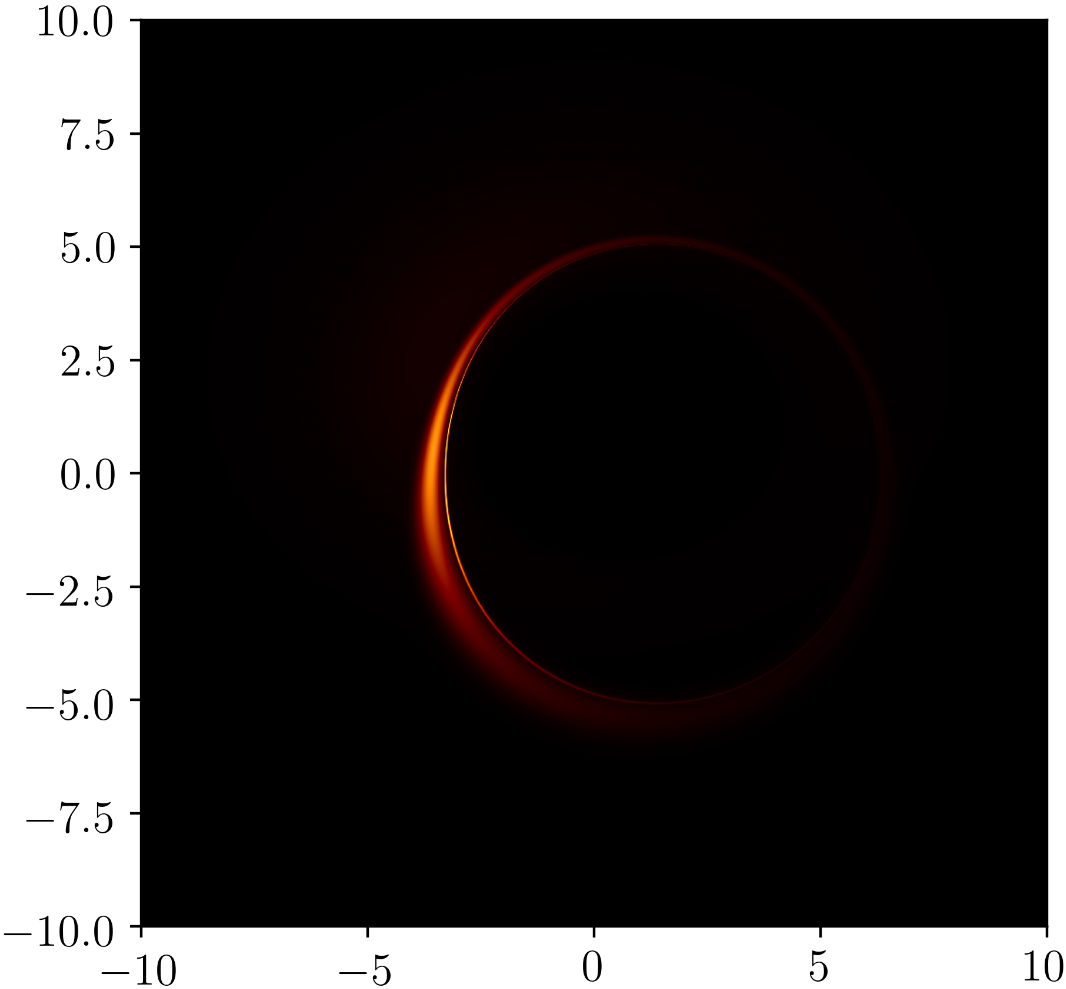}
    \includegraphics[width=2.1in, height=2.1in]{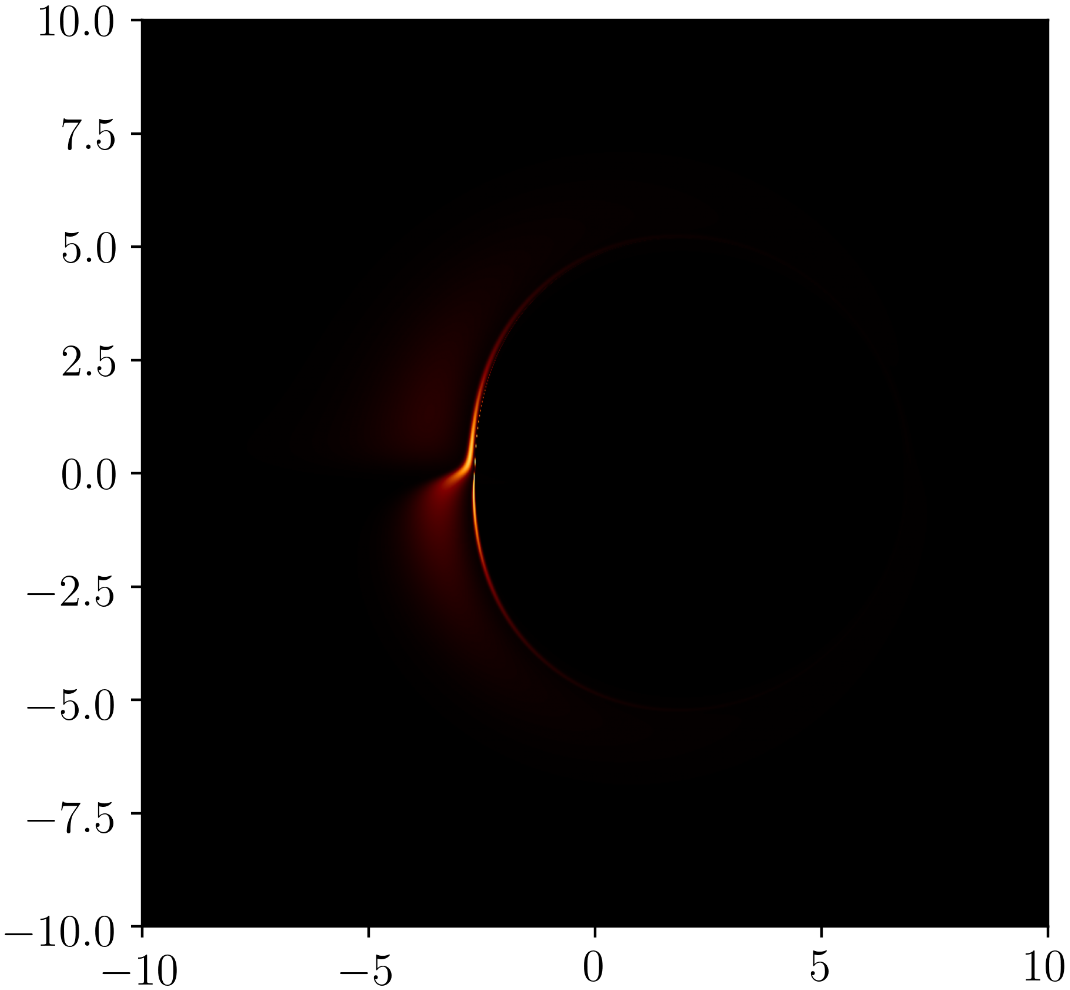}\\
    \textbf{\hspace{1.5cm} \large $\alpha(M)$}
    \caption{Image pixels \((\alpha, \beta)\) produced for different values of spin parameters (the top, middle and bottom row indicate a\(=[0,0.3,0.94]\), respectively) and inclination angles $i = 17^\circ, 45^\circ, 85^\circ$.}
    \label{fig:shadows}
\end{figure*}
In this section, we present our analysis on black-hole shadows and visibility amplitudes using M87$^*$ as a reference.  We use the \texttt{AART} code~\cite{low_lum_2024,Desire_2025} to trace the photon's trajectory backward in time. Accordingly, we need to define fiducial values for some parameters of the code, such as M87$^*$ distance, $d_M = 5.215 \times 10^{23}$m. It defines a mass parameter $\psi$ that is set to $1.075$, where the inverse, $1/\psi$, corresponds to $6.2 \times 10^9$ in solar masses. Using the Newtonian gravitational constant $G = 6.67 \times 10^{-11} \, \text{m}^3/\text{kg s}^2$, the speed of light $c = 2.99792458 \times 10^8 \, \text{m/s}$, and the solar mass $M_\odot = 1.988435 \times 10^{30} \, \text{kg}$, one has the mass of M87$^*$ in kilograms as $M_{\text{kg}} = 6.2 \times 10^9 \times \psi \times M_\odot$. This is then converted into a gravitational radius (meters) by $M_{\text{M87$^*$}} = M_{\text{kg}} \cdot G / c^2$. To convert angular size from radians to microarcseconds, we define $\mu\text{as\_to\_rad} = \pi / 648000 \times 10^{-6}$. Then, the unit conversion factor becomes $\text{unitfact} = 1 / (\mu\text{as\_to\_rad} \times 10^9)$. Moreover, the angular size of the black-hole in microarcseconds is calculated as $\theta = \arctan(M_{\text{M87$^*$}} / d_M) / \mu\text{as\_to\_rad}$.

Given a black-hole with spin \( a \) and an observer inclination angle \( \theta_o \), we can define a celestial coordinate system in the observer's sky using the coordinates \( (\alpha, \beta) \). These coordinates correspond to the apparent position of a light ray in the observer's image plane, where \(\alpha(M)\) represents the horizontal displacement of a photon in the image, primarily influenced by its angular momentum and \(\beta(M)\) represents the vertical displacement, which is influenced by the photon's motion perpendicular to the equatorial plane.  For each pixel in the image screen, corresponding to a direction \( (\alpha, \beta) \). This means computing the full geodesic equation to determine where and how the photon traveled through the curved spacetime around the black-hole with emission ring for three inclination views, \( i = 17^\circ \) (nearly face-on), \( i = 45^\circ \) (intermediate view) and \( i = 85^\circ \) (nearly edge-on). 

We present our results in Figure (\ref{fig:shadows}) that shows a $3\times 3$ grid images of the black-hole shadows for inclination angles $i = 17^\circ, 45^\circ, 85^\circ$. The spin values are set in the top row as $a=0$, the middle row with $a=0.3$, and the bottom row with $a=0.94$. At low inclination ($i = 17^\circ$), we have a symmetric nearly circular ring with minimal asymmetry similar to expected face-on appearance. At moderate inclination ($i = 45^\circ$), we can identify a clear Doppler beaming brightness that shifts to the left/right side (depending on rotation direction). The shadow appears slightly compressed but vertically consistent with frame dragging. At high Inclination ($i = 85^\circ$), we have a strong asymmetry that shifts left/right brightness contrast that is maximum at extreme spin a=0.94. As spin grows, we have that the limb-brightening becomes more prominent and the appearance of lens flare-like caustics indicate the influence of complex geodesics in the embedded space.

For a comparison with \text{M87$^*$}, our simulations predict a scale that remains similar across all inclinations, aligning with the EHT's observed shadow of about 40$\mu$\text{as} in diameter, which is consistent with expectations from a Kerr black-hole. At intermediate and high inclinations, both simulations and the EHT results show asymmetry: our model reveals clear brightness shifts, while the EHT is observed a bright southern arc. Regarding ring structure, our simulations consistently produce a distinct photon ring at all inclinations, comparable to the circular ring with variable brightness reported by the EHT results.    
\vphantom{}
\begin{figure*}[t!]
    \centering
     \rotatebox{90}{\textbf{$\vphantom{\beta(M)}$}}
    \includegraphics[width=2.15in, height=2.15in]{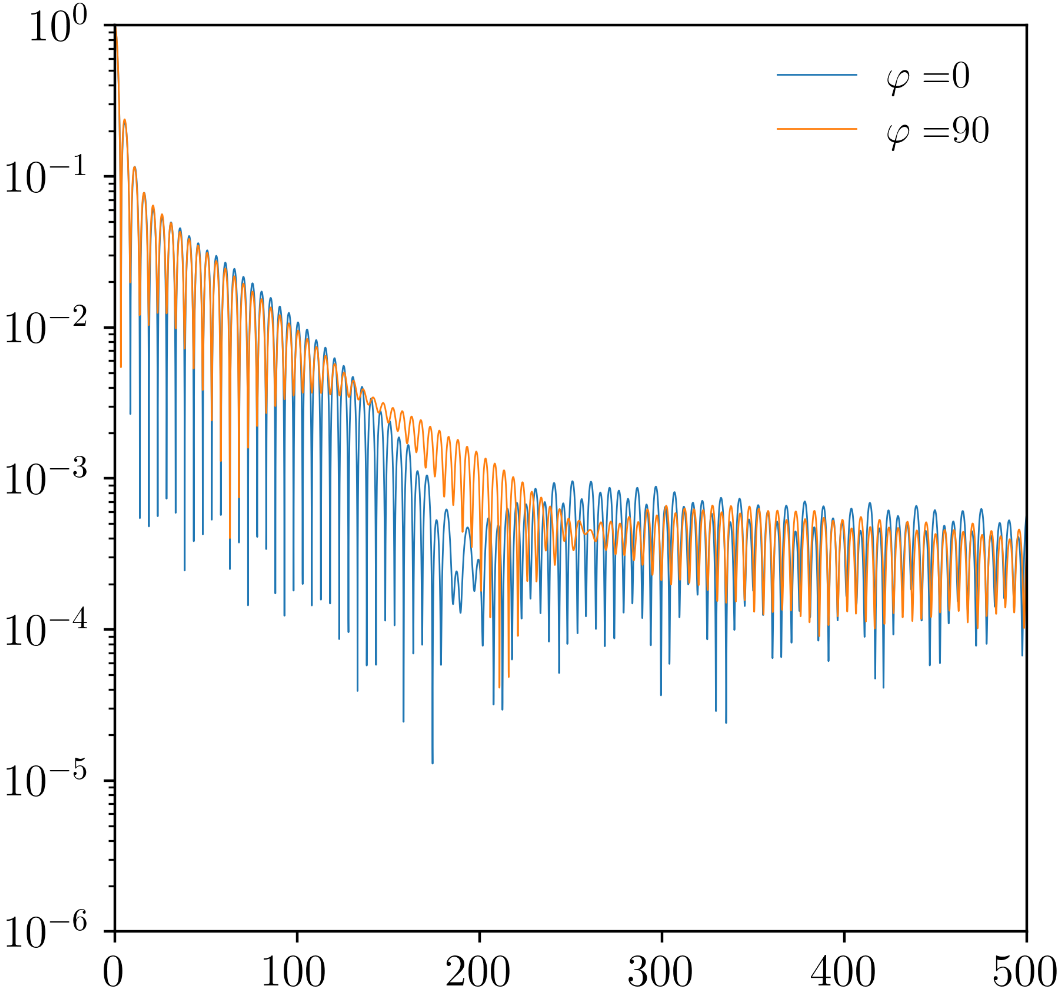}
    \includegraphics[width=2.15in, height=2.15in]{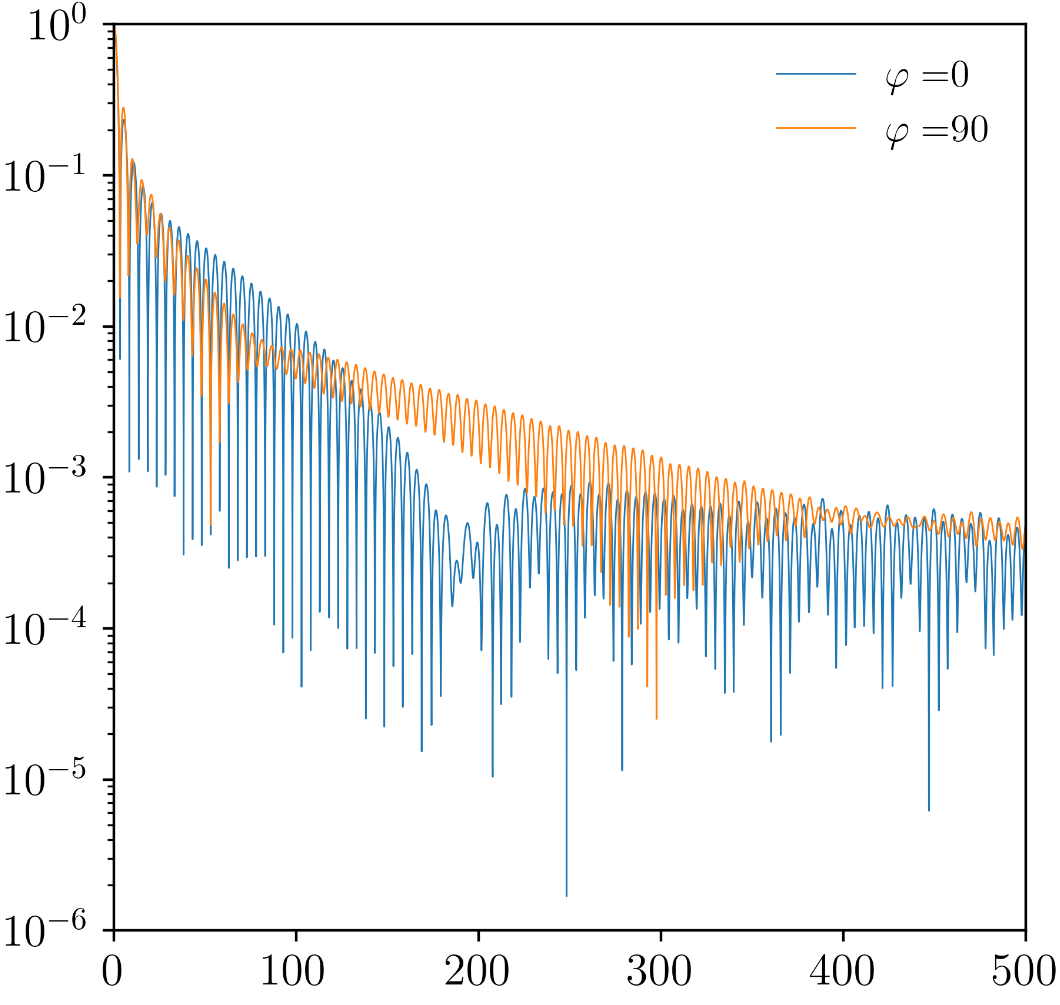}
    \includegraphics[width=2.15in, height=2.15in]{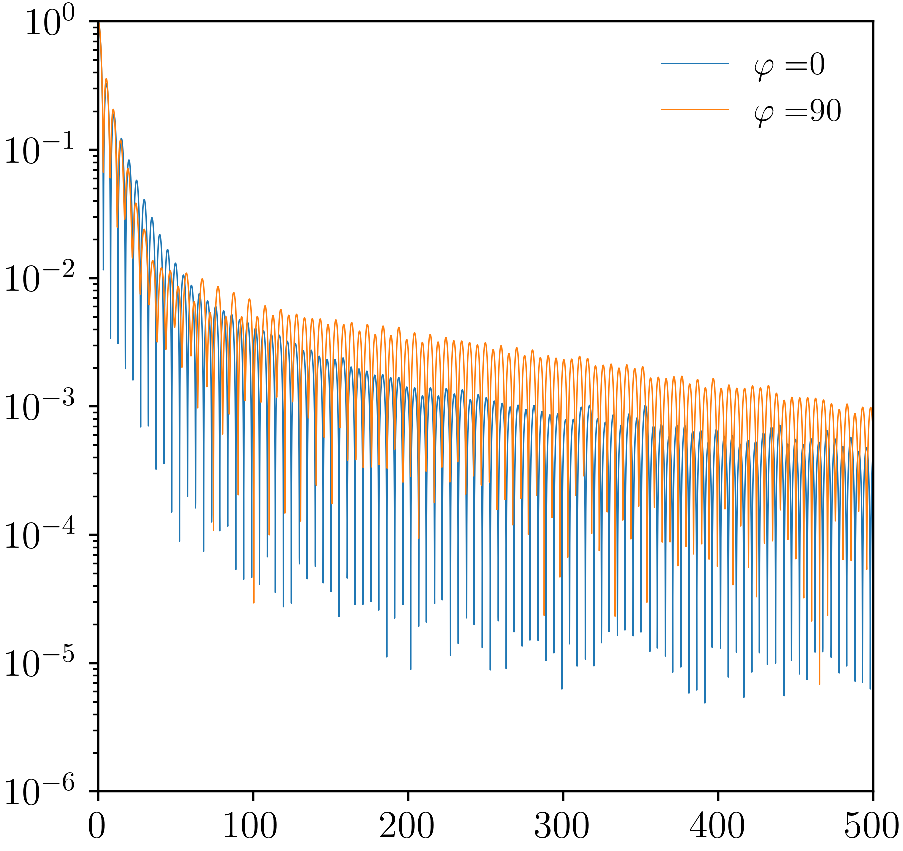} \\
    \rotatebox{90}{\textbf{Visibility Amplitude (Jy)\vspace{1cm}}}
    \includegraphics[width=2.15in, height=2.15in]{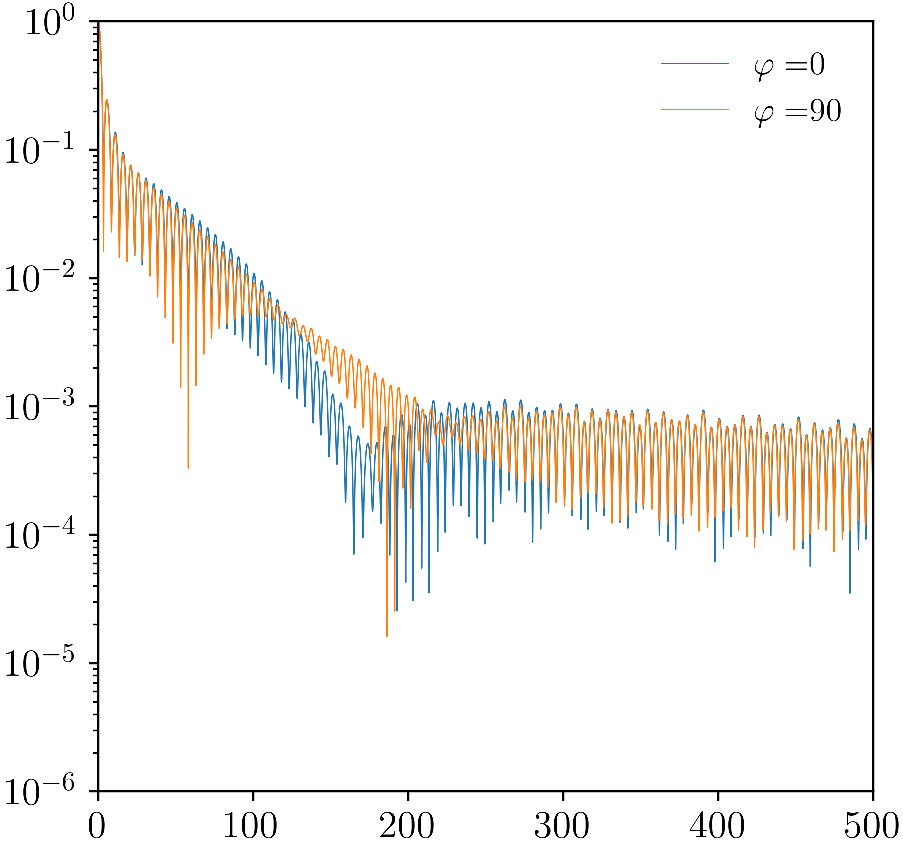}
    \includegraphics[width=2.15in, height=2.15in]{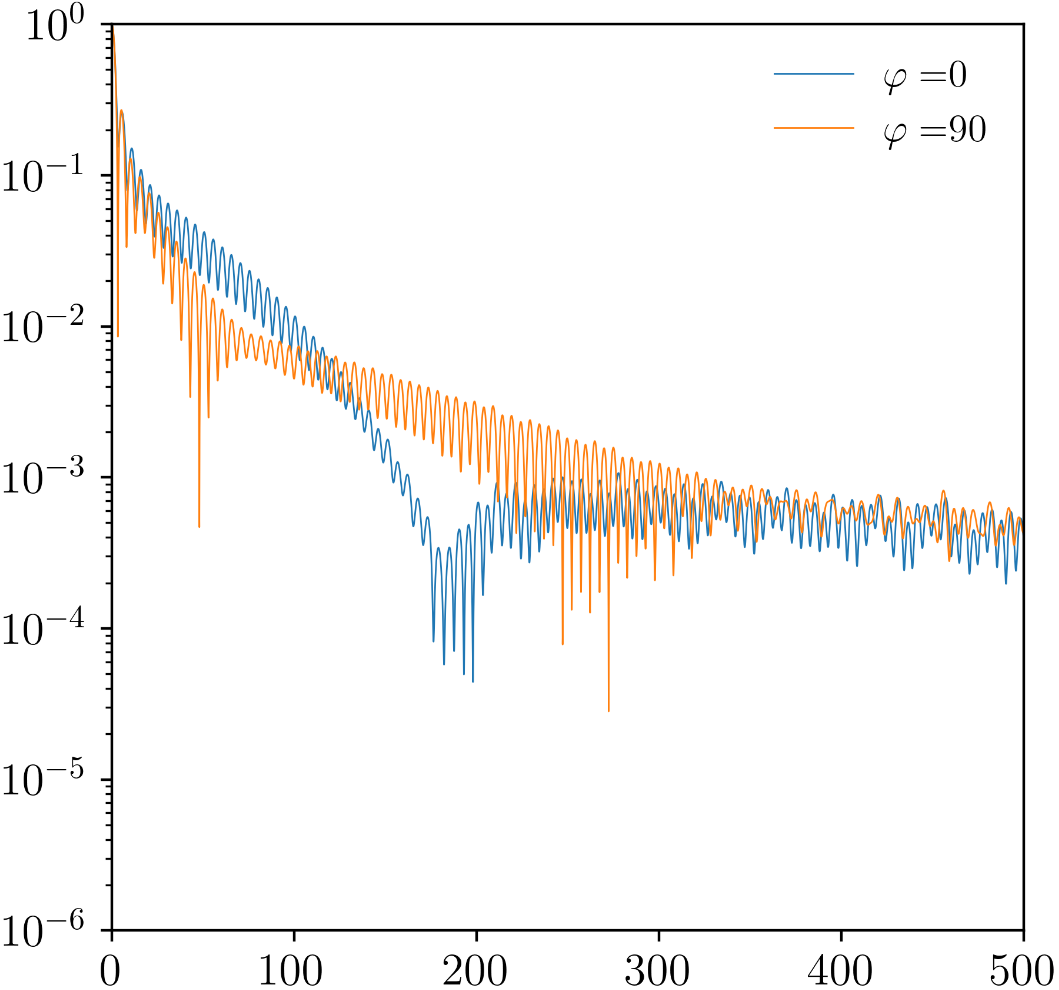}
    \includegraphics[width=2.15in, height=2.15in]{vis_0_90_a_0.01_i_85.png} \\
    \rotatebox{90}{\textbf{$\vphantom{\beta(M)}$}}
    \includegraphics[width=2.15in, height=2.15in]{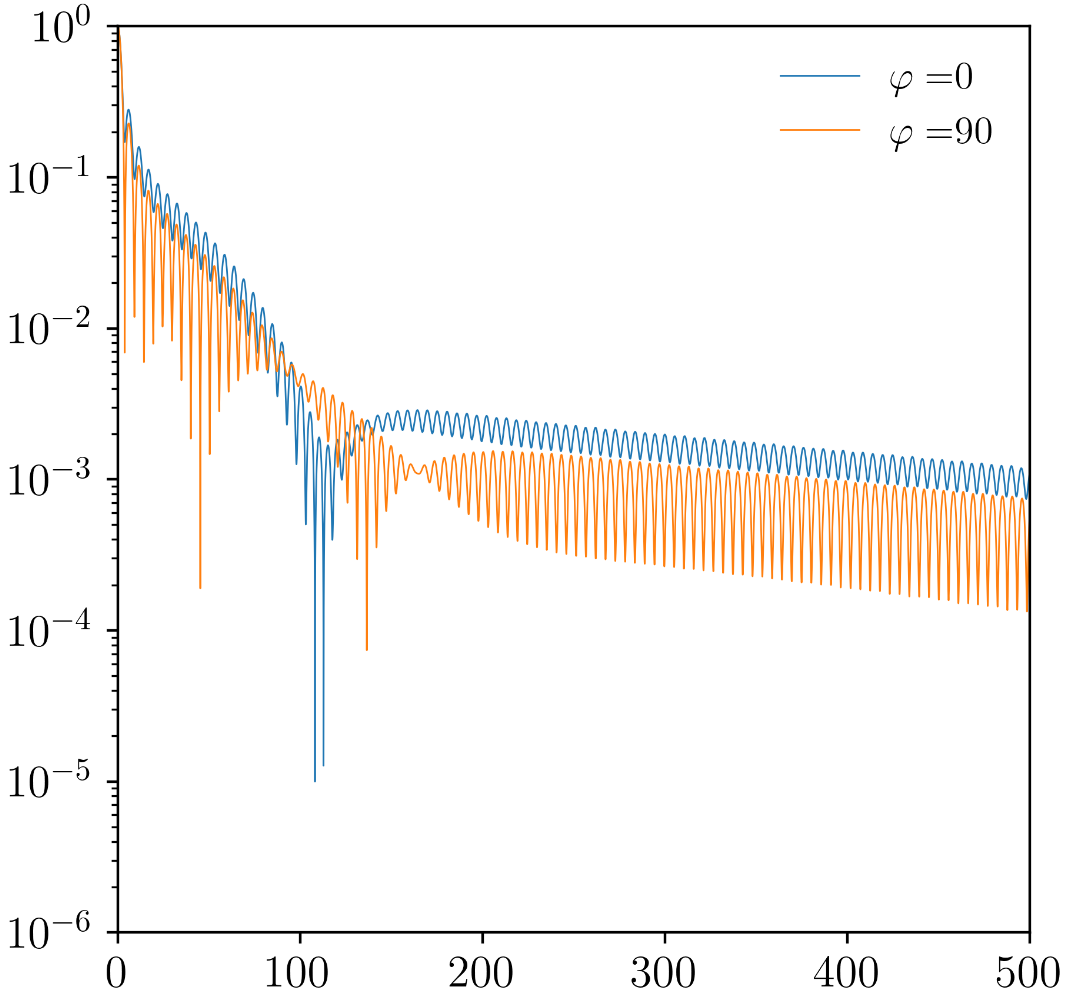}
    \includegraphics[width=2.15in, height=2.15in]{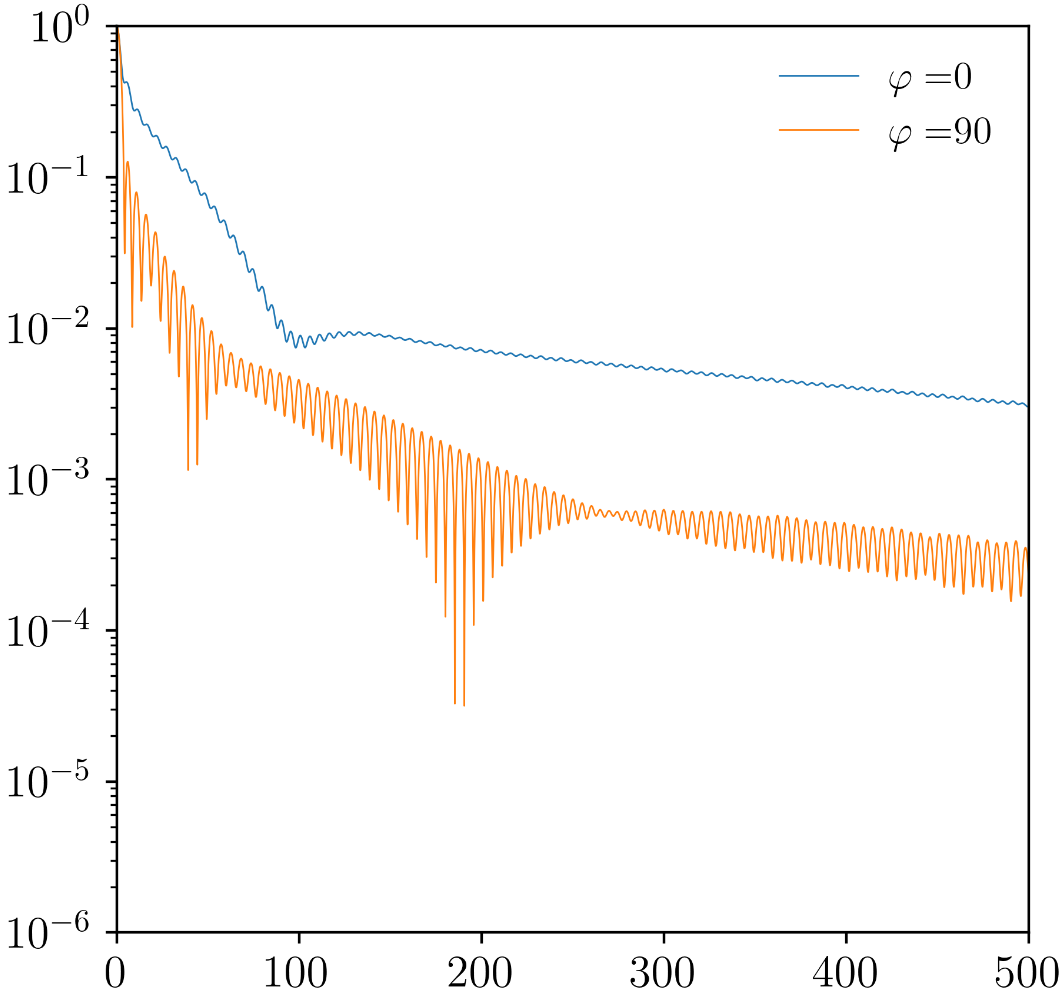}
    \includegraphics[width=2.15in, height=2.15in]{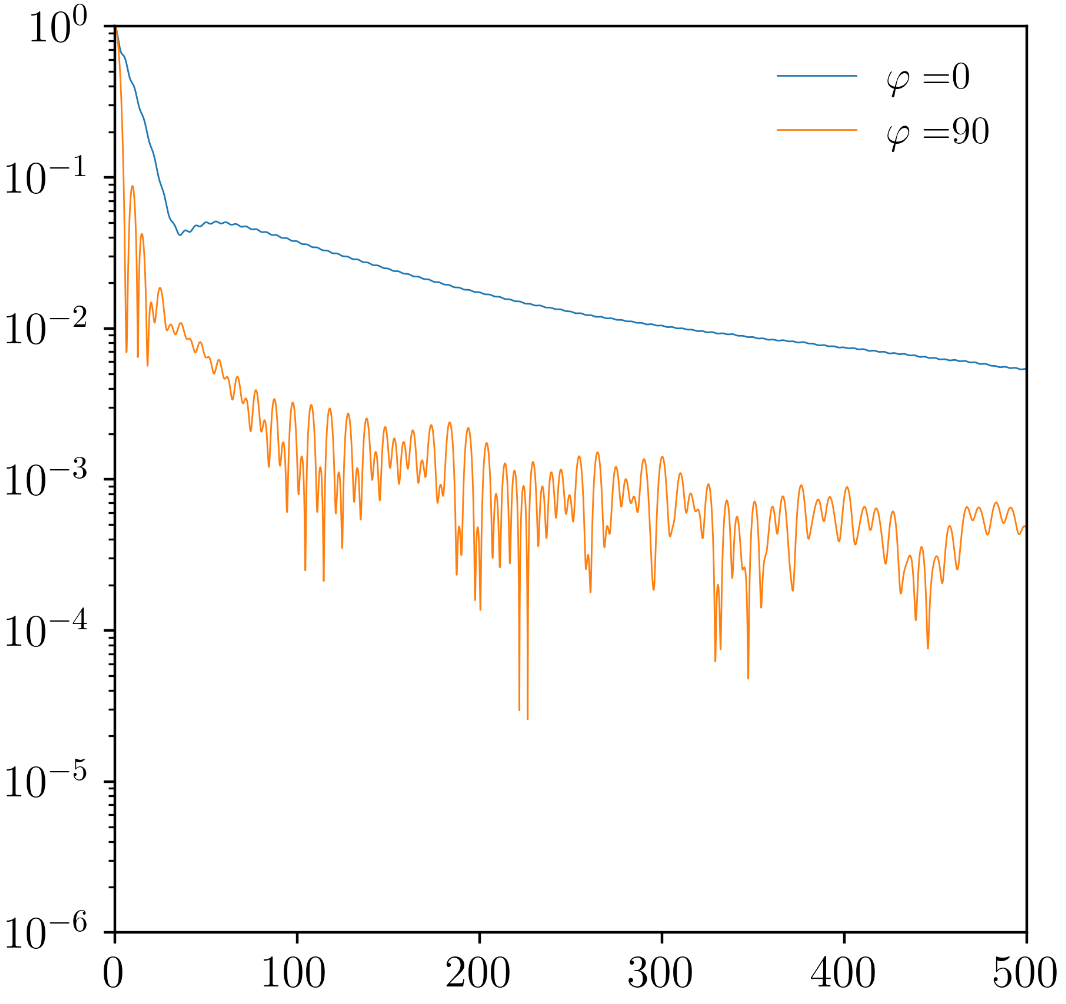}\\
    \textbf{\hspace{1.5cm} Baseline length (G$\lambda$)}
    \caption{Schematic showing visibility amplitude as a function of baseline length for two different azimuthal angles, \( \varphi = 0^\circ \) and \( \varphi = 90^\circ \) at different inclinations, (\( i = 17^\circ, 45^\circ, 85^\circ \)). From top to bottom rows, we have the spin parameters \(a=0,0.3,0.94\), respectively}
    \label{fig:visibilityampl}
\end{figure*}

In Figure (\ref{fig:visibilityampl}), we show our $3\times 3$ grid of plots of visibility amplitude (in Jy) as a function of baseline length (in Giga-lambda, GA), for different azimuthal angles $\varphi$ and polar angles $i$. Basically, these curves represent Fourier transforms of the black-hole shadow images used in Very Long Baseline Interferometry (VLBI), like in the EHT observations of M87$^{\ast}$. All plots show a quasi-periodic fringe structure, more prominent in some orientations. This reflects the sharp ring-like features of the shadow. We also point out that the visibility amplitude decreases rapidly with baseline, which indicates a compact source with fine structure, and it is also sensitive to azimuthal angle $\varphi$, especially at high inclination ($i \sim 85^\circ$). Moreover, some orientations show asymmetry, likely due to Doppler boosting and lens-thickening effects, which affect the brightness distribution across the shadow. The central baseline region has the highest amplitude that is expected due to the bright central structure (accretion flow or photon ring). As a comparison, M87$^{\ast}$ shows a similar behavior in EHT data, with nulls and visibility minima as well.

Our results show that spin also governs the depth of oscillations in the visibility amplitude. For instance, we see in Figure (\ref{fig:visibilityampl}) the upper left corner that has deeper, more symmetric oscillations due to lower spin. Moving right or downward, we see that the nulls become less sharp, and oscillations appear more chaotic or weaker. For high spins and large inclinations, visibility amplitudes vary significantly with baseline orientation, reflecting the asymmetric brightness distribution introduced by rotation. This effect is particularly strong when the line of sight is aligned with the boosted side of the accretion flow. In visibility grids, such spin effects manifest as a progression from symmetric and coherent oscillations at low spin and inclination, to weaker, less regular oscillations with pronounced azimuthal dependence at higher spin values. Thus, as spin increases, symmetry is broken of the azimuthal angles, especially at high $i$, leading to modulated visibility. The first null shifts slightly, and higher-order lobes lose coherence as well.

These theoretical predictions are consistent with the observational results obtained by the EHT for M87$^{\ast}$. The data revealed visibility nulls near $60 G\lambda$, corresponding to a ring diameter of approximately $42 \mu$as, in agreement with a Kerr black-hole of spin $a/M \sim 0.5-0.94$ viewed at an inclination of $\sim 17^{\circ}$. The Doppler-boosted brightness distribution observed in the image domain was also consistent with moderate spin and low inclination. More importantly, the EHT analysis demonstrated that visibility oscillations and shadow morphology provide complementary spin diagnostics, reinforcing our results.

When applied to alternative metrics such as the Gürses–Gürsey five-dimensional black-hole, these diagnostic tools remain valuable. Our preliminary analysis indicates that the Gürses–Gürsey metric preserves many Kerr-like features in how spin affects shadow morphology and visibility structure, though subtle deviations may exist. All in all, at high spin and low inclination, the resulting shadow and visibility characteristics remain compatible with the EHT observations of M87$^{\ast}$. This suggests that current VLBI observations cannot easily distinguish between the standard Kerr metric and certain higher-dimensional extensions, although future data with improved baseline coverage and dynamic range may be able to resolve such differences.

%
%



\section{Final remarks}
Our work builds on Nash's embedding theorem, which extends the geometry of spacetime into a higher-dimensional bulk. In contrast with traditional braneworld models--such as the Randall–Sundrum framework and its variants--our approach treats the extrinsic curvature as a genuine dynamical degree of freedom, rather than a mere consequence of matter fields. This enables a covariant, model-independent framework in which spacetime deformations can give rise to richer gravitational dynamics. The embedding formalism naturally introduces corrections to the gravitational field, influencing black hole horizons, geodesics, shadow morphology, and even cosmological evolution.
To explore these effects, we investigate an extension of the Kerr spacetime using the Gürses–Gürsey metric embedded in a five-dimensional bulk. This construction gives rise to a new parameter,  $\alpha_0$, inherited from cosmological perturbations. Acting as an effective tidal charge, $\alpha_0$
 serves as a relic of the extrinsic geometry. It modifies the conditions for horizon formation by slightly lowering the extremal spin threshold compared to the standard Kerr solution, thereby making horizon formation more restrictive. Meanwhile, the innermost stable circular orbit (ISCO) retains Kerr-like behavior, shrinking inward with increasing spin. These changes impact the structure and radiative efficiency of accretion disks, and may influence observable features such as brightness profiles and shadow size—positioning the ISCO as a critical probe for distinguishing standard Kerr BHs from their higher-dimensional embedded counterparts.

We also discuss the nonlocal nature of the extrinsic curvature contributions. In terms of Komar mass, a positive $\alpha_0$ reduces the effective mass at small $r$. This suggests a stronger attractive field at small scales. Then, the parameter $\alpha_0$ plays the role of an effective tidal charge, analogous to the projected bulk Weyl curvature in traditional braneworld models. By introducing an effective energy–momentum tensor (EMT), we analyzed the associated energy conditions and found that the strong energy condition (SEC) is violated. This violation implies that the effective matter induces repulsive gravitational effects, which can counteract gravitational collapse. As a result, the geometry becomes a viable candidate for singularity avoidance—potentially describing regular BHs, wormhole throats, or, in cosmological settings, phenomena such as dark energy and inflationary dynamics. The marginal satisfaction of the weak (WEC), null (NEC), and dominant energy conditions (DEC) further suggests that the source does not correspond to ordinary baryonic matter or classical field configurations. Instead, it may be interpreted as a projection of higher-dimensional geometric effects onto the four-dimensional spacetime. These features justify interpreting the embedded solution as a regularized Kerr-type geometry with effective dark-energy-like stress, meriting further study through linear perturbation and quasinormal mode analysis.

Using the \texttt{AART} ray-tracing code, simulations of shadow images and visibility amplitudes were produced for M87$^{\ast}$. As compared with EHT, we obtained shadow diameter $\sim 40 \mu$as that matches the Kerr prediction. Asymmetric ring brightness and visibility nulls are also consistent with EHT's bright southern arc. Furthermore, our simulations reveal that both spin and inclination exert a strong influence on shadow morphology and visibility amplitude oscillations. The spin parameter, in particular, imprints distinct and measurable signatures on these observables. For example, the spacing between visibility nulls encodes the shadow diameter, the depth of the nulls reflects the degree of brightness asymmetry, and azimuthal variations trace spin-induced Doppler beaming effects. These spin-dependent features have already been leveraged by the Event Horizon Telescope (EHT) to constrain the properties of M87$^{\ast}$, and they offer considerable potential for probing the nature of astrophysical BHs with future high-resolution VLBI observations.

\begin{acknowledgements}
AJSC acknowledges Conselho Nacional de Desenvolvimento Cient\'{i}fico e Tecnologico (CNPq) for the partial financial support for this work (Grant No. 305881/2022-1) and Fundação da Universidade Federal do Paraná (FUNPAR, Paraná Federal University Foundation) through public notice 04/2023-Pesquisa/PRPPG/UFPR for the partial financial support (Process No. 23075.019406/2023-92), and AJSC and CHC-A are very thankful for the financial support of the NAPI ``Fenômenos Extremos do Universo" of Fundação de Apoio à Ciência, Tecnologia e Inovação do Paraná, (NAPI FÍSICA – FASE 2), under protocol No 22.687.035-0. CHC-A also acknowledges the financial support of Fundação de Apoio à Ciência, Tecnologia e Inovação do Paraná (Grant No. PRD2023361000304).
\end{acknowledgements}

\bibliographystyle{spphys} 
\bibliography{sn-bibliography}

\begin{thebibliography}{10}
\providecommand{\url}[1]{{#1}}
\providecommand{\urlprefix}{URL }
\expandafter\ifx\csname urlstyle\endcsname\relax
  \providecommand{\doi}[1]{DOI \discretionary{}{}{}#1}\else
  \providecommand{\doi}{DOI \discretionary{}{}{}\begingroup \urlstyle{rm}\Url}\fi

\bibitem{batal}
A.~El-Batal, J.~Miller, M.~Reynolds, S.~Boggs, F.~Chistensen, W.~Craig, F.~Fuerst, C.~Hailey, F.~Harrison, D.~Stern, et~al., The Astrophysical Journal Letters \textbf{826}(1), L12 (2016)

\bibitem{bambi2013rotating}
C.~Bambi, L.~Modesto, Physics Letters B \textbf{721}(4-5), 329 (2013)

\bibitem{reynolds2021observational}
C.S. Reynolds, Annual Review of Astronomy and Astrophysics \textbf{59}(1), 117 (2021)

\bibitem{kerr}
R.P. Kerr, Physical Review Letters \textbf{11}(5), 237 (1963)

\bibitem{akiyama}
K.~Akiyama, A.~Alberdi, W.~Alef, J.C. Algaba, R.~Anantua, K.~Asada, R.~Azulay, U.~Bach, A.K. Baczko, D.~Ball, et~al., The Astrophysical Journal Letters \textbf{930}(2), L12 (2022)

\bibitem{kocherlakota}
P.~Kocherlakota, L.~Rezzolla, H.~Falcke, C.M. Fromm, M.~Kramer, Y.~Mizuno, A.~Nathanail, H.~Olivares, Z.~Younsi, K.~Akiyama, et~al., Physical Review D \textbf{103}(10), 104047 (2021)

\bibitem{Nash1956}
J.~Nash, Annals of Mathematics \textbf{63}(1), 20 (1956).
\newblock \doi{10.2307/1969989}.
\newblock \urlprefix\url{https://www.jstor.org/stable/1969989}

\bibitem{Randall1999}
L.~Randall, R.~Sundrum, Phys. Rev. Lett. \textbf{83}, 3370 (1999).
\newblock \doi{10.1103/PhysRevLett.83.3370}.
\newblock \urlprefix\url{https://journals.aps.org/prl/abstract/10.1103/PhysRevLett.83.3370}

\bibitem{Randall1999b}
L.~Randall, R.~Sundrum, Physical Review Letters \textbf{83}(23), 4690 (1999).
\newblock \doi{10.1103/PhysRevLett.83.4690}.
\newblock \urlprefix\url{https://journals.aps.org/prl/abstract/10.1103/PhysRevLett.83.4690}

\bibitem{ArkaniHamed1998}
N.~Arkani-Hamed, S.~Dimopoulos, G.~Dvali, Phys. Lett. B \textbf{429}, 263 (1998).
\newblock \doi{10.1016/S0370-2693(98)00466-3}.
\newblock \urlprefix\url{https://www.sciencedirect.com/science/article/pii/S0370269398004663}

\bibitem{Dvali2000}
G.~Dvali, G.~Gabadadze, M.~Porrati, Phys. Lett. B \textbf{485}, 208 (2000).
\newblock \doi{10.1016/S0370-2693(00)00669-9}.
\newblock \urlprefix\url{https://www.sciencedirect.com/science/article/pii/S0370269300006699}

\bibitem{Battye2001}
R.~Battye, B.~Carter, Physics Letters B \textbf{509}(3-4), 331 (2001).
\newblock \doi{10.1016/S0370-2693(01)00495-6}.
\newblock \urlprefix\url{https://arxiv.org/abs/hep-th/0101061}

\bibitem{Davis2003}
S.C. Davis, Physical Review D \textbf{67}(2), 024030 (2003).
\newblock \doi{10.1103/PhysRevD.67.024030}.
\newblock \urlprefix\url{https://arxiv.org/abs/hep-th/0208205}

\bibitem{Yamauchi2007}
D.~Yamauchi, M.~Sasaki, Progress of Theoretical Physics \textbf{118}(2), 245 (2007).
\newblock \doi{10.1143/PTP.118.245}.
\newblock \urlprefix\url{https://academic.oup.com/ptp/article/118/2/245/1910064}

\bibitem{Perlick2021}
V.~Perlick, O.Y. Tsupko, Living Reviews in Relativity \textbf{24}(1), 3 (2021).
\newblock \doi{10.1007/s41114-021-00032-6}.
\newblock \urlprefix\url{https://arxiv.org/abs/2105.07101}

\bibitem{Akiyama2019}
K.~Akiyama, others (Event Horizon Telescope~Collaboration), The Astrophysical Journal Letters \textbf{875}(1), L1 (2019).
\newblock \doi{10.3847/2041-8213/ab0ec7}.
\newblock \urlprefix\url{https://doi.org/10.3847/2041-8213/ab0ec7}

\bibitem{Medeiros2016}
L.~Medeiros, C.k. Chan, F.~{\"O}zel, D.e.a. Psaltis, arXiv preprint  (2016).
\newblock \urlprefix\url{https://arxiv.org/abs/1601.06799}

\bibitem{Johnson2020}
M.D. Johnson, R.~Narayan, D.~Psaltis, L.~Medeiros, et~al., Science Advances \textbf{6}(12), eaaz1310 (2020).
\newblock \doi{10.1126/sciadv.aaz1310}.
\newblock \urlprefix\url{https://www.science.org/doi/10.1126/sciadv.aaz1310}

\bibitem{Chael2016}
A.A. Chael, M.D. Johnson, R.~Narayan, S.S. Doeleman, J.F.C. Wardle, The Astrophysical Journal \textbf{829}(1), 11 (2016).
\newblock \doi{10.3847/0004-637X/829/1/11}.
\newblock \urlprefix\url{https://arxiv.org/abs/1601.06173}

\bibitem{Amarilla2012}
L.~Amarilla, E.F. Eiroa, Physical Review D \textbf{85}(6), 064019 (2012).
\newblock \doi{10.1103/PhysRevD.85.064019}.
\newblock \urlprefix\url{https://journals.aps.org/prd/abstract/10.1103/PhysRevD.85.064019}

\bibitem{Vagnozzi2022}
S.~Vagnozzi, L.~Visinelli, P.~Lüftinger, E.~Barausse, B.~Afrin, Classical and Quantum Gravity \textbf{39}(7), 075007 (2022).
\newblock \doi{10.1088/1361-6382/ac5f3c}.
\newblock \urlprefix\url{https://arxiv.org/abs/2205.07787}

\bibitem{Nedkova2013}
P.G. Nedkova, V.K. Tinchev, S.S. Yazadjiev, Physical Review D \textbf{88}(12), 124019 (2013).
\newblock \doi{10.1103/PhysRevD.88.124019}.
\newblock \urlprefix\url{https://journals.aps.org/prd/abstract/10.1103/PhysRevD.88.124019}

\bibitem{Akiyama2025}
K.~Akiyama, others (Event Horizon Telescope~Collaboration), Astronomy \& Astrophysics \textbf{693}, A265 (2025).
\newblock \doi{10.1051/0004-6361/202451296}.
\newblock \urlprefix\url{https://www.aanda.org/articles/aa/full_html/2025/04/aa51296-24/aa51296-24.html}

\bibitem{Wielgus2022}
M.~Wielgus, others (Event Horizon Telescope~Collaboration), The Astrophysical Journal Letters \textbf{930}(1), L12 (2022).
\newblock \doi{10.3847/2041-8213/ac6429}.
\newblock \urlprefix\url{https://arxiv.org/abs/2205.04623}

\bibitem{Mizuno2018}
Y.~Mizuno, Z.~Younsi, C.M. Fromm, O.~Porth, H.~Olivares, L.~Rezzolla, M.~Moscibrodzka, H.~Falcke, M.~Kramer, Nature Astronomy \textbf{2}, 585 (2018).
\newblock \doi{10.1038/s41550-018-0449-5}.
\newblock \urlprefix\url{https://www.nature.com/articles/s41550-018-0449-5}

\bibitem{Roelofs2019}
F.~Roelofs, H.~Falcke, C.~Brinkerink, M.~Kramer, S.~Markoff, Z.~Younsi, Y.~Mizuno, Astronomy \& Astrophysics \textbf{625}, A124 (2019).
\newblock \doi{10.1051/0004-6361/201732423}.
\newblock \urlprefix\url{https://www.aanda.org/articles/aa/full_html/2019/05/aa32423-17/aa32423-17.html}

\bibitem{gursey}
M.~Gürses, F.~Gürsey, Journal of High Energy Physics \textbf{16}, 2385–2390 (1975).
\newblock \doi{10.1063/1.522480}.
\newblock \urlprefix\url{https://doi.org/10.1063/1.522480}

\bibitem{Capistrano2024}
A.J.S. Capistrano, R.C. Nunes, L.A. Cabral, Physical Review D \textbf{109}(12), 123517 (2024).
\newblock \doi{10.1103/PhysRevD.109.123517}.
\newblock \urlprefix\url{https://journals.aps.org/prd/abstract/10.1103/PhysRevD.109.123517}

\bibitem{low_lum_2024}
A.~C\'ardenas-Avenda\~no, A.~Lupsasca, H.~Zhu, Phys. Rev. D \textbf{107}, 043030 (2023).
\newblock \doi{10.1103/PhysRevD.107.043030}.
\newblock \urlprefix\url{https://link.aps.org/doi/10.1103/PhysRevD.107.043030}

\bibitem{Desire_2025}
T.~Desire, A.~Cárdenas-Avendaño, A.~Chael, The Astrophysical Journal \textbf{980}(2), 262 (2025).
\newblock \doi{10.3847/1538-4357/adac4d}.
\newblock \urlprefix\url{https://dx.doi.org/10.3847/1538-4357/adac4d}

\bibitem{greene}
R.E. Greene, Memoirs of the American Mathematical Society \textbf{97} (1970).
\newblock \urlprefix\url{https://www.ams.org/books/memo/0097/}

\bibitem{MAIA20029}
M.~Maia, E.M. Monte, Physics Letters A \textbf{297}(1), 9 (2002).
\newblock \doi{https://doi.org/10.1016/S0375-9601(02)00182-2}.
\newblock \urlprefix\url{https://www.sciencedirect.com/science/article/pii/S0375960102001822}

\bibitem{GDE}
M.D. Maia, E.M. Monte, J.M.F. Maia, J.S. Alcaniz, Classical and Quantum Gravity \textbf{22}(9), 1623 (2005).
\newblock \doi{10.1088/0264-9381/22/9/010}.
\newblock \urlprefix\url{https://dx.doi.org/10.1088/0264-9381/22/9/010}

\bibitem{Maia_2007}
M.D. Maia, N.~Silva, M.C.B. Fernandes, Journal of High Energy Physics \textbf{2007}(04), 047 (2007).
\newblock \doi{10.1088/1126-6708/2007/04/047}.
\newblock \urlprefix\url{https://dx.doi.org/10.1088/1126-6708/2007/04/047}

\bibitem{gde2}
M.D. Maia, A.J.S. Capistrano, J.S. Alcaniz, E.M. Monte, Gen. Rel. Grav. \textbf{43}, 2685 (2011).
\newblock \doi{10.1007/s10714-011-1192-8}

\bibitem{Capistrano2021}
A.J.S. Capistrano, Physics of the Dark Universe \textbf{33}, 100872 (2021).
\newblock \doi{10.1016/j.dark.2021.100872}.
\newblock \urlprefix\url{https://doi.org/10.1016/j.dark.2021.100872}

\bibitem{capistrano2022}
A.J.S. Capistrano, L.A. Cabral, J.A.P.F. Marão, C.H. Coimbra-Araújo, The European Physical Journal C \textbf{82}, 1434 (2022).
\newblock \doi{10.1140/epjc/s10052-022-10431-9}

\bibitem{Capistrano2024b}
A.J.S. Capistrano, C.H. Coimbra-Ara{\'u}jo, R.~de~C{\'a}ssia~dos Anjos, Universe \textbf{10}(9), 355 (2024).
\newblock \doi{10.3390/universe10090355}.
\newblock \urlprefix\url{https://www.mdpi.com/2218-1997/10/9/355}.
\newblock Published 4 September 2024; open access

\bibitem{Carron_2022}
J.~Carron, M.~Mirmelstein, A.~Lewis, Journal of Cosmology and Astroparticle Physics \textbf{2022}(09), 039 (2022).
\newblock \doi{10.1088/1475-7516/2022/09/039}.
\newblock \urlprefix\url{https://dx.doi.org/10.1088/1475-7516/2022/09/039}

\bibitem{Rosenberg:2022sdy}
E.~Rosenberg, S.~Gratton, G.~Efstathiou, Mon. Not. Roy. Astron. Soc. \textbf{517}(3), 4620 (2022).
\newblock \doi{10.1093/mnras/stac2744}

\bibitem{BICEPKeck}
P.A.R. Ade, et~al., Phys. Rev. Lett. \textbf{127}, 151301 (2021).
\newblock \doi{10.1103/PhysRevLett.127.151301}.
\newblock \urlprefix\url{https://link.aps.org/doi/10.1103/PhysRevLett.127.151301}

\bibitem{6dFGalaxy}
F.~Beutler, C.~Blake, M.~Colless, D.H. Jones, L.~Staveley-Smith, L.~Campbell, Q.~Parker, W.~Saunders, F.~Watson, Monthly Notices of the Royal Astronomical Society \textbf{416}(4), 3017 (2011).
\newblock \doi{10.1111/j.1365-2966.2011.19250.x}.
\newblock \urlprefix\url{https://doi.org/10.1111/j.1365-2966.2011.19250.x}

\bibitem{Ross:2014qpa}
A.J. Ross, L.~Samushia, C.~Howlett, W.J. Percival, A.~Burden, M.~Manera, Mon. Not. Roy. Astron. Soc. \textbf{449}(1), 835 (2015).
\newblock \doi{10.1093/mnras/stv154}

\bibitem{Alam:2016hwk}
S.~Alam, et~al., Mon. Not. Roy. Astron. Soc. \textbf{470}(3), 2617 (2017).
\newblock \doi{10.1093/mnras/stx721}

\bibitem{mgcamb2023}
Z.~Wang, S.H. Mirpoorian, L.~Pogosian, A.~Silvestri, G.B. Zhao, Journal of Cosmology and Astroparticle Physics \textbf{2023}(08), 038 (2023).
\newblock \doi{10.1088/1475-7516/2023/08/038}.
\newblock \urlprefix\url{https://dx.doi.org/10.1088/1475-7516/2023/08/038}

\bibitem{cobaya}
J.~Torrado, A.~Lewis, Journal of Cosmology and Astroparticle Physics \textbf{2021}(05), 057 (2021).
\newblock \doi{10.1088/1475-7516/2021/05/057}.
\newblock \urlprefix\url{https://dx.doi.org/10.1088/1475-7516/2021/05/057}

\bibitem{DESI2024VI}
{DESI Collaboration}, A.G. Adame, et~al., JCAP \textbf{02}, 021 (2025).
\newblock \doi{10.1088/1475-7516/2025/02/021}.
\newblock \urlprefix\url{https://arxiv.org/abs/2404.03002}

\bibitem{BlandfordZnajek1977}
R.D. Blandford, R.L. Znajek, Monthly Notices of the Royal Astronomical Society \textbf{179}, 433 (1977).
\newblock \doi{10.1093/mnras/179.3.433}.
\newblock \urlprefix\url{https://doi.org/10.1093/mnras/179.3.433}

\bibitem{Dadhich2000}
N.~Dadhich, R.~Maartens, P.~Papadopoulos, V.~Rezania, Physics Letters B \textbf{487}(1–2), 1–6 (2000).
\newblock \doi{10.1016/S0370-2693(00)00798-X}.
\newblock \urlprefix\url{https://doi.org/10.1016/S0370-2693(00)00798-X}

\end{thebibliography}

\end{document}